\documentclass[%
 reprint,
nofootinbib,
 amsmath,amssymb,
 aps,
floatfix,
]{revtex4-1}

\usepackage{xspace}
\usepackage{xcolor}
\usepackage[normalem]{ulem}
\usepackage[breaklinks, plainpages=false, colorlinks=true, anchorcolor=cyan, linkcolor=magenta, citecolor=cyan, urlcolor=purple, bookmarks=false]{hyperref}
\usepackage{appendix}
\usepackage{multirow}
\usepackage{graphicx}
\usepackage{acronym}
\usepackage{bm}

\newacro{imbh}[IMBH]{intermediate-mass black hole}
\newacro{bhns}[BHNS]{black hole neutron star}
\newacro{bbh}[BBH]{binary black hole}
\newacro{bh}[BH]{black hole}
\newacro{bns}[BNS]{binary neutron star}
\acrodef{FAR}[FAR]{false alarm rate}
\newacro{bf}[BF]{Bayes' factor}
\newacro{cbc}[CBC]{compact binary coalescence}
\newacro{ce}[CE]{Cosmic Explorer}
\acrodef{SNe}[SNe]{Supernova}
\newacro{da}[DA]{data analysis}
\newacro{et}[ET]{Einstein Telescope}
\newacro{eob}[EOB]{Effective-One-Body}
\newacro{fd}[FD]{frequency domain}
\newacro{gw}[GW]{gravitational wave}
\newacro{gr}[GR]{General relativity}
\newacro{hm}[HM]{Higher mode}
\newacro{ifo}[IFO]{interferometer}
\newacro{imr}[IMR]{inspiral-merger-ringdown}
\newacro{im}[IM]{inspiral-to-merger}
\newacro{kagra}[KAGRA]{Kamioka Gravitational Wave Detector}
\newacro{ligo}[LIGO]{Laser Interferometer Gravitational-Wave Observatory}
\newacro{lso}[LSO]{Last Stable Orbit}
\newacro{lvc}[LVC]{LIGO-Virgo Collaboration}
\newacro{lvk}[LVK]{LIGO-Virgo-Kagra Collaboration}
\newacro{lo}[LO]{leading order}
\newacro{ns}[NS]{neutron star}
\newacro{nr}[NR]{numerical relativity}
\newacro{pn}[PN]{post-Newtonian}
\newacro{pe}[PE]{parameter estimation}
\newacro{psd}[PSD]{power spectral density}
\newacro{asd}[ASD]{amplitude spectral density}
\acrodef{KN}[KN]{kilonova}
\newacro{xg}[XG]{next-generation}
\newacro{jsd}[JSD]{jensen shannon divergence}
\newacro{qnm}[QNM]{quasi-normal mode}

\acrodefplural{KN}[KNe]{kilonovae}
\newacro{qc}[QC]{quasi-circular}
\newacro{snr}[SNR]{signal-to-noise ratio}
\acrodef{SNR}[SNR]{signal-to-noise ratio}
\newacro{ng}[NG]{Next Generation}
\newacro{eos}[EoS]{Equation of State}

\newcommand{\nrsur}{NRSur7dq4\xspace}

\newcommand{\dynesty}{\texttt{dynesty}\xspace}
\newcommand{\tstartgeo}{\Delta t^\mathrm{geo}_{\mathrm{start}}}

\newcommand{\fundamental}{\mathrm{Kerr}_{220}}
\newcommand{\overtone}{\mathrm{Kerr}_{221}}

\newcommand{\DSone}{\mathrm{DS}_1}
\newcommand{\DStwo}{\mathrm{DS}_2}

\begin{document}

\title{Black-hole ringdown analysis with inspiral-merger informed templates and limitations of classical spectroscopy}

\author{Koustav Chandra$^{1,2}$}
\email{kbc5795@psu.edu}
\author{Juan Calder\'on Bustillo$^{3,4}$}
\email{juan.calderon.bustillo@gmail.com}

\affiliation{$^{1}$Institute for Gravitation and the Cosmos, Department of Physics and Department of Astronomy and Astrophysics, The Pennsylvania State University, University Park, Pennsylvania 16802, USA}
\affiliation{$^{2}$Max Planck Institute for Gravitational Physics (Albert Einstein Institute), Am M\"uhlenberg 1, 14476 Potsdam, Germany}
\affiliation{$^{3}$Instituto Galego de F\'isica de Altas Enerx\'ias, Universidade de Santiago de Compostela, 15782 Santiago de Compostela, Galicia, Spain}
\affiliation{$^{4}$Department of Physics, The Chinese University of Hong Kong, Shatin, New Territories, Hong Kong}

\begin{abstract}

The ringdown of a perturbed black hole consists of a superposition of quasi-normal modes (QNMs), with complex frequencies determined by the black hole's mass and spin, while phases and amplitudes depend on binary parameters. Traditional semi-agnostic ringdown analyses ignore the remnant's binary-merger origin, assuming instead generic perturbations. This can lead to an unphysical number of free parameters and usage of priors inconsistent with those imposed by full inspiral-merger-ringdown (IMR) analysis. Here we revisit an alternative approach, first proposed in \citet{CalderonBustillo:2020rmh}, in which ringdowns are modeled using the post-merger portion of full IMR waveform models. This naturally includes all QNMs without adding extra degrees of freedom and ties priors to binary parameters. We analyze the signal GW150914 with the post-merger portion of the IMR surrogate model NRSur7dq4, finding decisive evidence for the no-hair theorem with a Bayes factor $>650:1$ or $>99.8$\% probability, compared to only $3:1$ from semi-agnostic spectroscopy, consistent with previous work. We also find modest evidence ($\simeq 5:1$) for the $(\ell,m,n)=(2,2,1)$ overtone and none for higher modes. Using simulated signals, we validate our formalism by showing accurate recovery of remnant properties. Next, by analysing simulated signals with post-merger signal-to-noise ratios up to 100, we show that classical spectroscopy based on overtone detection alone cannot provide strong no-hair tests. Finally, we demonstrate that the non-orthogonality of QNMs makes the inferred mode content strongly model-dependent, leading to multiple inconsistent interpretations that are equally well-supported by the data. We briefly discuss potential implications for recent events, such as GW231123 and GW250114.

\end{abstract}

\maketitle

\section{Introduction} \label{sec:intro}

\Ac{gr}, predicts that a perturbed astrophysical \ac{bh} transitions to a stationary Kerr \ac{bh} by emitting \acp{gw} via a process known as ringdown~\citep{Vishveshwara:1970cc, Teukolsky:1973ha, Chandrasekhar:1975zza}. This emission consists of a spectrum of \acp{qnm} which, according to the no-hair theorem, are characterised by complex frequencies uniquely determined by the \ac{bh}'s mass and spin~\citep{Carter:1971zc, Leaver:1985ax} through well-known relations~\citep{Berti:2006wq}. In contrast, the amplitudes and phases of the \acp{qnm} are determined by the nature of the perturbation, which, in the case of quasi-spherical \acp{bbh}, is given by the component masses and spins through a significantly less understood map~\citep{Kamaretsos:2012bs, Kamaretsos:2011um, London:2014cma, DelPozzo:2016kmd, London:2018gaq, Borhanian:2019kxt, Forteza2021, Carullo2024, Gennari:2023gmx}. 

The observation of \acp{gw} from \ac{bbh} mergers by the Advanced LIGO and Virgo detectors~\citep{LIGOScientific:2014pky, VIRGO:2014yos} has provided a unique opportunity to study \acp{qnm}, leading to the development of numerous methods to detect them~\citep{Brito:2018rfr, Carullo:2019flw, CalderonBustillo:2020rmh,  Ghosh:2021mrv, Bhagwat:2021kfa, Capano:2021etf, Isi:2021iql, Maggio:2022hre, Ma:2022wpv, Ma:2023vvr, Pacilio:2024qcq, Pompili:2025cdc,  Crescimbeni:2024sam, Lu:2025mwp}. These have made it possible to study the \ac{gw} emission from the merger remnant, enabling novel tests of \ac{gr} in the strong-field dynamical regime. Notable among these studies are tests of the no-hair theorem, which search for potential deviations from the spectrum predicted by \ac{gr}, as well as \acf{imr} consistency tests, which assess whether the properties of the final \ac{bh} are consistent with \ac{gr} predictions based on the binary's initial parameters~\citep{LIGOScientific:2016lio, Ghosh:2017gfp, Dhanpal:2018ufk, LIGOScientific:2019fpa, LIGOScientific:2020tif, LIGOScientific:2021sio}.

This type of study, however, has several challenges. First, the exponentially damped nature of \acp{qnm} causes the \ac{snr} to fall off rapidly after the merger, suggesting that \acp{qnm} should be searched for immediately after it. Such an approach was historically disfavoured due to the expected presence of non-linearities right after the merger, which may pollute the analysis. This caused original analyses to start approximately $10M$ (where $M$ denotes the remnant's mass in seconds) after the merger, so that a linear \ac{qnm}-based description of the signal is valid~\citep{LIGOScientific:2016lio}. However, \citet{Giesler:2019uxc} showed that including overtones along with the fundamental $(\ell = 2, m=2, n=0)$ \ac{qnm} can provide a good description of the \ac{gw} signal at times close to the merger, claiming a smooth transition from inspiral to merger and ringdown. Subsequently,~\citet{Isi:2019aib} claimed evidence for the presence of the $(\ell=2, m=2, n=1)$ overtone in the signal GW150914~\citep{LIGOScientific:2016aoc}, using this to test the no-hair theorem to the $20\%$ level. Such a claim, however, was later contested by~\citet{Cotesta:2022pci, Correia:2023bfn} and remains unresolved.

It is worth noting that even before the work of~\citet{Giesler:2019uxc}, it was understood that overtones yielded a better fit to the \ac{gw} signal at the merger~\citep{London:2014cma, Baibhav:2017jhs}. However, this does not necessarily imply that the spacetime is accurately described by a linearly perturbed \ac{bh}. In fact,~\citet{Cheung:2022rbm, Mitman:2022qdl} showed that non-linear (quadratic) \acp{qnm} are excited in \ac{nr} simulations of quasi-spherical \ac{bbh} mergers. 

Independent of the true nature or actual detection of overtones, \citet{CalderonBustillo:2020rmh} showed that tests of the no-hair theorem based on their detection could not provide strong evidence in favour of the Kerr hypothesis even in the case of extremely loud signals where the ringdown has an \ac{snr} of 100, almost an order of magnitude larger than in GW150914. In fact, a similar test performed on the recent event GW250114 -- shown in Fig. 4 of ~\citep{LIGOScientific:2025epi, LIGOScientific:2025obp} -- could only provide a probability of $\sim 3:1$ in favour of the Kerr nature of the remnant, qualitatively similar to the result on GW150914. This is due to the increasing number of free parameters needed to fit the ever-increasing number of individually resolvable overtones for increasing \ac{snr}, which induces a large Ockham penalty on such models. This motivated the authors to propose the use of ringdown models whose \ac{qnm} phases and amplitudes are restricted to those allowed by the quasi-spherical origin of the final \ac{bh}. Because such models only exist for reduced portions of the parameter space and for a reduced number of \acp{qnm}~\citep{DelPozzo:2016kmd, Forteza2021, Gennari:2023gmx, Carullo2024}, the authors proposed using the post-merger portion of complete \ac{imr} models as a ringdown model. On the one hand, these naturally include the whole set of \ac{qnm} modes, including overtones and nonlinear modes. On the other hand, the phases and amplitudes of such modes are all constrained to those allowed by the nature of the progenitor binary, highly reducing the Ockham penalty paid by the model and allowing for the imposition of consistent priors in \ac{imr} consistency tests. Using such a model, the authors provided a $99.7\%$ confirmation of the Kerr nature of the GW150914 remnant object.

In this work, we revisit and improve upon the approach proposed in \citet{CalderonBustillo:2020rmh} in two main ways. First, in that work, the authors employed the phenomenological waveform model IMRPhenomPv2~\citep{Khan:2018fmp}, which is restricted to the quadrupolar $(2, \pm 2)$ mode. This is, its ringdown portion only includes the fundamental $(2,2,0)$ quasi-normal mode and the corresponding infinite tower of overtones $(2,2,n > 0)$. In contrast, we now permit the use of any waveform model. In particular, in this work, we make use of the \nrsur waveform model \citep{Varma:2019csw}, which is directly trained on \ac{nr} simulations of non-eccentric precessing \acp{bbh} waveforms~\citep{Boyle:2019kee}. This model includes higher-order $(\ell,m) \neq (2,\pm 2)$ modes, up to $\ell \leq 4$. Therefore, its ringdown portion naturally includes the corresponding $(\ell,m,0)$ \acp{qnm} together with the associated $(\ell,m,n > 0)$ overtones. Importantly, because the model is ``calibrated'' to \ac{nr} waveforms, it self-consistently incorporates the phases, amplitudes, and mode asymmetries of the \acp{qnm}.

An alternative, similar in spirit, is provided by \ac{nr}-informed post-merger models~\citep{DelPozzo:2016kmd, Gennari:2023gmx}. Here, the ringdown is modeled as a superposition of the longest-lived ($n=0$) \acp{qnm}, with amplitudes and phases calibrated to \ac{nr} simulations of \ac{bbh} mergers. Essentially, they are the post-peak portion of \ac{imr} waveform TEOBResumS-Dalí in its non-eccentric, non-precessing limit~\citep{Damour:2014yha, Nagar2018, Nagar:2020pcj}. Overtones and nonlinearities are not included explicitly but are incorporated phenomenologically through time-dependent amplitude parameters, ensuring a smooth transition to the preceding inspiral waveform. In addition, these models are currently limited to non-precessing and non-eccentric binaries, although progress is being made to obtain extensions to, e.g., eccentric cases \cite{Carullo2024}. By contrast, by simply using the post-peak portion of full \ac{imr} models (like \nrsur\ ), we make sure that all relevant \acp{qnm}, including overtones, and non-linearities are included by construction, providing a more faithful description of the post-peak signal. Moreover, our approach is agnostic to the specific waveform family, as it can be used with any existing model and extended to any future models, avoiding the step of individually calibrating each of the modes.

Second, ringdown studies require the isolation of the post-merger signal to avoid the analysis being polluted by the inspiral phase. This leads to a sharply starting signal that requires either a time-domain analysis that avoids spectral leakage \cite{Carullo:2019flw} and contamination from untargeted portions of the data or, in principle, a modified frequency-domain analysis that considers the noise correlations introduced by the abruptly starting signal \cite{Talbot2021}. Instead, \citet{CalderonBustillo:2020rmh} opted to replace the pre-ringdown part of GW150914 with real glitch-free noise around the signal to perform a frequency-domain analysis using the standard data power spectrum. While they checked that the results were consistent with those obtained on simulated signals, such a method cannot generally be considered stable, as it naturally leads to a signal containing a discontinuity at the point where the noise is blended with the ringdown data. Instead, this work implements a frequency-domain analysis based on the inpainting of the ``irrelevant'' portion of the data and signal following the work of~\citet{Capano:2021etf}. 

In this work, we apply our ringdown analysis technique to GW150914 and validate our analysis on simulated signals consistent with this event. We first demonstrate that our method yields accurate estimates of the final mass and spin. Next, we analyze the GW150914 ringdown using both classical semi-agnostic approaches and our \ac{bbh}-informed ringdown model. With the semi-agnostic analyses, we reproduce previous findings: marginal evidence for the $(2,2,1)$ overtone together with no support for higher-order angular modes with $(\ell,m) \neq (2,\pm 2)$. Finally, in agreement with~\citet{CalderonBustillo:2020rmh}, we show that our \ac{bbh}-informed model is strongly favored over a ``hairy'' model given by free damped sinusoids with a Bayes factor exceeding 650, thereby confirming the Kerr nature of the remnant to sub-percent level.

The rest of this paper is organised as follows. We begin by summarizing the frequency-domain method adopted in this article in Section~\ref{sec:paint}. Next, we describe the three waveform models considered in this study in Section~\ref{sec:waveform}. Section~\ref{sec:analysis-setup} summarises our setups for the analysis of GW150914 and that of a synthetic signal consistent with it. In Section~\ref{sec:results}, we first report our results for GW150914 under different models. We estimate the remnant properties of GW150914, search for evidence for the presence of the first overtone of the fundamental ringdown mode, and test the no-hair theorem. Further, we discuss the range of applicability of overtone models based on the consistency between mode-amplitude estimates at a given time and those inferred from analyses at different times. We find that the overtone model produces self-consistent results after \(\tstartgeo=0M\), where $\tstartgeo$ is the geocentric time, in units of the remnant black-hole mass, measured after the strain peak. We also observe that it correctly predicts the overtone amplitude until \(\tstartgeo=3M\), which coincides with the instant at which both the overtone and single-mode models fit the data equally well, while the model restricted to the fundamental mode can only achieve self-consistency after $\tstartgeo=3M$. Next, we focus on agnostic mode recovery, discussing the physical interpretation of our results. In particular, we highlight how such a purely agnostic analysis based on free damped sinusoids yields three main ways to reproduce the GW150914 signal. We argue that while one of such avenues is consistent with the fundamental mode plus overtone paradigm, our results demonstrate that (semi-)agnostic mode recovery is rather ill-defined due to the non-orthogonality of quasi-normal modes. Next in Section~\ref{sec:gw150914-like}, we confirm our results using a synthetic signal consistent with GW150914 and discuss the level to which current methods -- as opposed to our \ac{bbh}-informed proposal -- enable strong tests of the no-hair theorem. We do the same for a high SNR signal in Section~\ref{sec:high-snr}. In these last two sections, we also show explicitly crucial limitations of classical \ac{bh} expectroscopy. In particular, we show how the inferred \ac{qnm} content of a given ringdown signal depends dramatically on model assumptions, even when the data equally statistically support such models. Finally, in Section~\ref{sec:conclusion}, we conclude our work with some final remarks.

\section{Painted Likelihood}\label{sec:paint}

Bayesian parameter estimation libraries, like Bilby~\citep{Ashton:2018jfp} or PyCBC-inference~\citep{Biwer:2018osg}, fundamentally assume that the discrete-time domain detector data $\boldsymbol{d}$ is a linear combination of a possible GW signal $\boldsymbol{s}(\boldsymbol{\theta})$ and instrument noise $\boldsymbol{n}$. In this work, we model \(\boldsymbol{s}\) using $\boldsymbol{h}_M(\boldsymbol{\theta})$ which is either (i) a \textit{hairy} model constructed as a superposition of damped sinusoids [Eq.~\eqref{eq:hair-model}], (ii) a Kerr waveform [Eq.~\eqref{eq:kerr}], or (iii) the post-inspiral portion of a \nrsur waveform~\citep{Varma:2019csw}.

The instrument noise $\boldsymbol{n}$, on the other hand, is typically assumed to be a set of random samples drawn from a wide-sense stationary, zero-mean Gaussian distribution. Such a distribution is fully described by the Covariance matrix $C_{ik} = \langle n_i n_k \rangle$, where $n_i$ and $n_k$ represent individual noise samples at discrete time instances $t_i$ and $t_k$, respectively.  

Imposing periodic boundary conditions leads to a diagonal Fourier-domain covariance matrix $\tilde{C}_{jj}$, with elements given by:
\begin{equation}
    \tilde{C}_{jj} = \langle |\tilde{n}_j^2 | \rangle = \frac{T}{2} S_j~.
\end{equation}
Here, $S_j$ is the one-sided noise power spectrum of the data and 
$$
    \tilde{n}_j= \sum_k n_k  e^{-2 \pi i j k f_s T}
$$
is the discrete Fourier transform of $\boldsymbol{n}$ at frequency bin index $j$ assuming that the data's sampling frequency is $f_s$ and duration is $T$. 

These assumptions significantly simplify the Bayesian computation, as, in this scenario, obtaining the single-detector log-likelihood function for the signal reduces to calculating the following noise-weighted inner product~\citep{Finn:1992wt, Veitch:2009hd}:
\begin{equation}\label{eq:signal-likelihood}
\begin{aligned}
\ln \mathcal{L}_S &= - \frac{\langle \boldsymbol{d} - \boldsymbol{h} \mid \boldsymbol{d} - \boldsymbol{h} \rangle}{2} + \mathrm{constant} \\
&= -\frac{2}{T} \sum_{j} \frac{|\tilde{d}_j - \tilde{h}_j|^2}{S_j} + \mathrm{constant}~.
\end{aligned}
\end{equation}
Unfortunately, post-inspiral analysis naturally involves the usage of abruptly-starting signals, which fail to meet the previous conditions. 

To eliminate the influence of the data (and the waveform) between some start time $t_\mathrm{start}$ and end time $t_\mathrm{end} > t_\mathrm{start}$ on $\ln \mathcal{L}_S$  we adopt the ``gating and inpainting'' method described in ~\citep{Zackay:2019kkv, Capano:2021etf}. This method assumes that the residual time-series $\boldsymbol{r}= \boldsymbol{d} - \boldsymbol{h}_M$ is made up of two additive components: residual $\boldsymbol{r}_g$ where the time between $t_\mathrm{start}$ and $t_\mathrm{end}$ is zeroed out and $\boldsymbol{x}$, a correction term that is zero at all times except within the gated time interval. To paint the gating time, we solve for 
$\boldsymbol{x}$ under the condition:
\begin{equation}
    \boldsymbol{C}^{-1}(\boldsymbol{x} + \boldsymbol{r}_g ) = 0
\end{equation}
within the gated time interval. We use the painted residual $\boldsymbol{r}_p = \boldsymbol{x} + \boldsymbol{d}_g - \boldsymbol{h}_g$ to define the modified frequency-domain likelihood:
\begin{equation}
    \ln \mathcal{L}_S' = - \frac{\langle \boldsymbol{r} \mid \boldsymbol{r}_p \rangle}{2} + \mathrm{constant}~.
\end{equation}
Our analysis uses this modified single-detector likelihood function and Bayes theorem to obtain the log posterior probability distribution of the signal parameters $\boldsymbol{\theta}$ :
\begin{equation}
    \ln p(\boldsymbol{\theta} \mid \boldsymbol{d}, M) = \ln \pi(\boldsymbol{\theta}\mid M) -\ln \mathcal{Z}_M + \sum_{k}^{N_\mathrm{IFO}} \ln \mathcal{L}_k'(\boldsymbol{d} \mid \boldsymbol{\theta}, M)~.
\end{equation}
Here, $\pi(\boldsymbol{\theta}\mid M)$ is the prior and $\mathcal{Z}_M$ is the model evidence defined as follows:
\begin{equation}
{\mathcal{Z}} = \int \prod_{k}^{N_\mathrm{IFO}} \pi(\boldsymbol{\theta} \mid M) {\mathcal{L}_k'}(\boldsymbol{d} |\boldsymbol{\theta}, M) d\boldsymbol{\theta}~,
\end{equation}
which amounts to the average of the likelihood over the prior space. Two key factors determine the above.
a) The model's ability to fit the data is reflected in the maximum likelihood value. b) Ockham’s razor effect, which penalises models with a large ``badly fitting'' volume of parameter space.

Given two different signal models $A$ and $B$, the ratio of the probabilities of the data given the model is given by the ratio
\begin{equation}
    \ln \mathrm{BF}^{A}_{B} = \ln \frac{\mathcal{Z}_A}{\mathcal{Z}_B}~,
\end{equation}

where $\rm BF^{A}_{B}$ is the relative Bayes Factor between the models A and B.

\section{Ringdown Models}\label{sec:waveform}

In this work, we consider three different ringdown models: the Hair model, the Kerr black-hole model, and the post-inspiral waveform of the \nrsur, which we refer to as the IM-R model. We model the GW strain $h_M(\boldsymbol{\theta})$~\footnote{We have adopted the non-vector notation here for convenience.} by projecting the waveform polarizations $h_{+,\times}$ onto the detector using the detector antenna patern $F_{+,\times}$ which depend on the source's sky-location $(\alpha,\delta)$ and polarisation angle $\psi$ as :
\begin{equation}
    \boldsymbol{h}_M(\boldsymbol{\theta}) = F_+h_+ + F_\times h_\times~.
\end{equation}
We perform all of our analyses using a fixed sky location and merger time. For the case of GW150914, we choose the maximum likelihood parameter values drawn from the event's publicly accessible posterior samples from~\citet{Islam:2023zzj}, effectively excluding uncertainties in their measurements. For our analysis of simulated signals, we fix the merger time and sky location to the true values. The remaining parameters $\boldsymbol{\theta}$ depend on the properties and orientation of the source, whose actual parametrization depends on our choice of model as we discuss next. 

\subsection{Hair model}
This model, commonly known as ``Damped Sinusoid'' model, represents the emitted ringdown signal as a superposition of free damped sinusoids (DSs) of the form
\begin{equation}\label{eq:hair-model}
    h_+ - ih_\times = \sum_{a=1}^{N} \mathcal{A}_a e^{i (\omega_a + i/\tau_a) (t-t_a)+\phi_a}~.
\end{equation}
The parameters ${\mathcal{A}_a, \omega_a, \tau_a, \phi_a}$ respectively denote the amplitude, central angular frequency, damping time, and initial phase of  $a^\mathrm{th}$ damped sinusoid. $t_a$ denotes the time elapsed after the peak of the inspiral-merger-ringdown waveform, which, depending on the analysis, is kept fixed.  This model allows all parameters to vary independently. In particular, damping times and frequencies are not tied to the properties of the final \ac{bh} and the corresponding amplitudes and phases are not characterised by the progenitor \ac{bbh}. So, a ``Hair'' model, \(\mathrm{DS}_N\), consisting of N modes has 4N degrees of freedom.

\subsection{Kerr model}

The Kerr waveform is a restricted version of the Hair model in which the complex mode frequencies $\Omega_{\ell m n} = \omega_{\ell m n} + i/\tau_{\ell m n}$ behave in accordance with the no-hair theorem. The mode takes the form
\begin{equation}\label{eq:kerr}
    h_+ - i h_\times = \sum_{\ell, m, n} 
    \mathcal{A}_{\ell m n}\,{}^{-2}S_{\ell m n}(\iota,\varphi,\chi_f)\,
    e^{i (t-t_a)\Omega_{\ell m n} + \phi_{\ell m n}}.
\end{equation}
where the parameters $\Omega_{\ell m n}$ are determined by the remnant’s redshifted mass $M_f$ and spin $\chi_f$. In addition, the angular dependence of the emission is captured by the spin-weighted spheroidal harmonics ${}^{-2}S_{\ell m n}$, which depend on the inclination $\iota$, azimuth $\varphi$, and $\chi_f$. The mode amplitudes $\mathcal{A}_{\ell m n}$ and phases $\phi_{\ell m n}$ are left as free parameters. An analysis including $N$ modes, therefore, introduces $2N+4$ free parameters. In the following, we respectively denote the Kerr models restricted to the fundamental mode and including $N$ overtones by \(\fundamental\) and \(\mathrm{Kerr}_{22N}\).

Here, we have assumed reflection symmetry of the remnant spacetime, which relates negative-$m$ modes to their positive-$m$ counterparts through complex conjugation. Under this assumption, the strain can be expressed entirely in terms of the positive-$m$ spectrum. We also neglect retrograde modes, whose excitation is suppressed for quasi-circular binaries and are not expected to be detectable at the \acp{snr} considered here.

\subsection{IM-R NRSur7dq4 model}

Our third model, which we denote by IM-R, restricts the \ac{qnm} amplitudes and phases to those consistent with perturbed black holes born from a quasi-spherical \ac{bbh} merger. This is achieved by using as a ringdown model the post-inspiral portion of a full \ac{imr} model such as \nrsur~\citep{Varma:2019csw}, which inherently contains all \acp{qnm} with phases and amplitudes fitted to numerical relativity simulations of \ac{bbh} mergers.

    \section{Analysis setup}\label{sec:analysis-setup}

    \subsection{Data}

We analyze publicly available strain data around GW150914, downsampled to $4096~\mathrm{Hz}$~\citep{LIGOScientific:2019lzm}. We apply a high-pass Butterworth filter with a cutoff frequency of $15~\mathrm{Hz}$ to suppress low-frequency noise. Given that our analysis is performed in the Fourier domain, we apply a Tukey window to mitigate spectral leakage.

For the likelihood evaluation, we set a minimum frequency of \( f_{\mathrm{min}} = 20~\mathrm{Hz} \), which we also use as the reference frequency \( f_{\mathrm{ref}} \) at which we quote spin-angle measurements in both the \ac{imr} and IM-R analyses. This is significantly below the peak amplitude frequency, \( f_{\mathrm{peak}} \sim 190~\mathrm{Hz} \) of the signal. For the data power spectrum estimate, we use 512 seconds of data around the event and use the median-mean Welch method~\cite{Allen:2005fk}.

We verify our results by analysing two different simulated signals injected in zero noise, using an Advanced LIGO Hanford and Livingston network operating at O1 sensitivity. The first simulated signal is a numerical relativity surrogate waveform consistent with the best-fit GW150914 waveform obtained by~\citet{Islam:2023zzj}. The second ccorresponds to the numerical relativity simulation \texttt{SXS:BBH:0305} with the same extrinsic parameters and total mass as the previous signal but rescaled in luminosity distance to achieve a ringdown network signal-to-noise ratio of $\sim 100$ and restricted to its quadrupole $(\ell,m) = (2,\pm 2)$ modes~\citep{Boyle:2019kee}.

    \subsection{Priors}

\paragraph{Hair analysis.} 
For the single damped sinusoid analysis, we adopt log-uniform priors on the amplitude $\mathcal{A}_0 \in (10^{-24}, 10^{-19})$, and uniform priors on the central frequency $f_0 \in (100,500)~\mathrm{Hz}$, damping time $\tau_0 \in (10^{-4},10^{-2})~\mathrm{s}$, and phase $\phi_0 \in (0,2\pi)$. In the case of multiple damped sinusoids, additional modes are included with $f_a \in (100,500)~\mathrm{Hz}$, $\tau_a \in (10^{-4},10^{-2})~\mathrm{s}$, and relative amplitude $\mathcal{A}_a \in (0,0.99)\prod_{k=0}^{a-1}\mathcal{A}_k$, which enforces a hierarchical amplitude structure.

\paragraph{Kerr analysis.} 
Here we impose uniform priors on the remnant mass $M_f \in (35,140)M_\odot$ and dimensionless spin $\chi_f \in (0,0.99)$, together with an isotropic prior on the inclination angle $\iota$. For the fundamental mode, we adopt a log-uniform prior on the amplitude $\mathcal{A}_{220} \in (10^{-24},10^{-19})$, while for overtones we use $\mathcal{A}_{22n}/\mathcal{A}_{220} \in (0,5)$ to allow them to exceed the fundamental. All phases are drawn uniformly from $(0,2\pi)$.

\paragraph{IMR and IM-R analysis} Consistent with \citet{Romero-Shaw:2020owr, LIGOScientific:2020ibl}, we adopt uniform priors on spin magnitudes (\(0 \leq \chi_{1,2} \leq 0.99\)) and detector-frame component masses \(m_i \in (10, 100)M_\odot\), assuming isotropic distributions for spin orientations, sky location, and binary orientation parameters (\(\iota, \phi\)). Due to the limitations of the \nrsur model, we constrain the mass ratio \(q\) between \(1/6 \leq q \leq 1\).

For the luminosity distance, we assume a uniform prior in the source frame, with \(\Lambda\)CDM cosmology with parameters \(H_0 = 67.9 \, \mathrm{km} \, \mathrm{s}^{-1} \, \mathrm{Mpc}^{-1}\) and \(\Omega_m = 0.3065\), as in \citet{LIGOScientific:2020ibl}, and $D_L$ constrained between 10 Mpc and 1 Gpc. 

    \subsection{Sampler settings}

We sample the parameter space using the \dynesty sampler in its static configuration, with 2000 live points~\citep{Speagle:2019ivv}. We employ random walks for sampling, with a minimum of 100 and a maximum of $10^{4}$ MCMC steps. The number of random walk steps in each chain is at least 50 times the autocorrelation length (\texttt{nact}). The run terminates once the remaining evidence fraction, \texttt{dlogz}, falls below 0.1, ensuring that the uncertainty in the evidence estimate is $\Delta \ln \mathcal{Z} \lesssim 0.1$.

    \section{Ringdown Analysis of GW150914}\label{sec:results}
\begin{figure}
    \centering
    \includegraphics[width=0.5\textwidth]{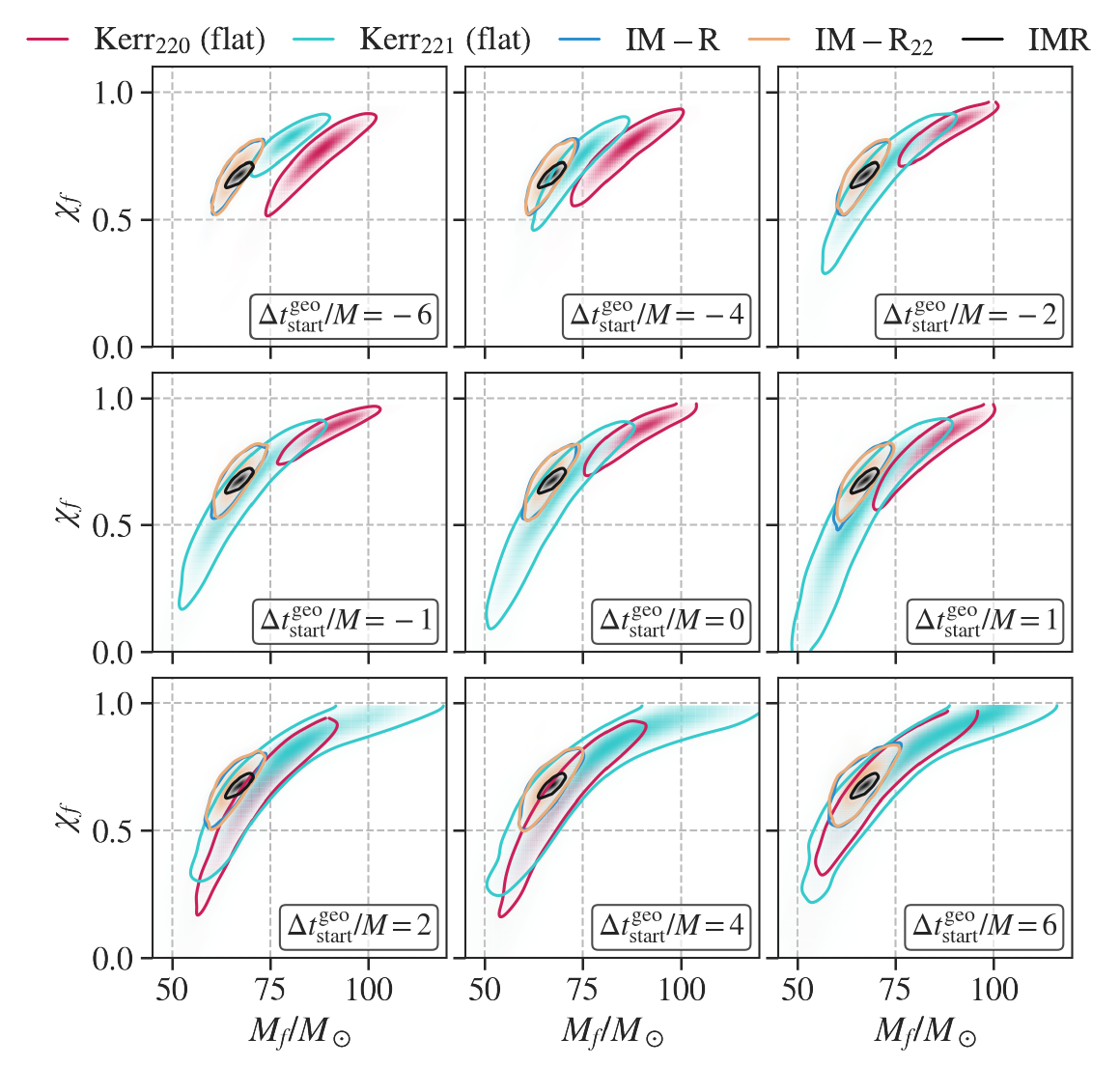}
    \caption{\textbf{Remnant mass and spin of GW150914 according to different ringdown models and from the full signal analysis}. Each panel shows results for a different analysis start time, $\tstartgeo$, measured from the signal peak in units of the final \ac{bh}'s mass. The contours enclose the 90\% credible regions.}
    \label{fig:corner}
\end{figure}

In this section, we present the results of our analysis of GW150914. We discuss measurements of the remnant mass and spin, as well as tests of the no-hair theorem. We explore different approaches, assess the presence of overtones and higher-order angular modes, and perform an agnostic reconstruction of the signal.  

    \subsection{Final mass and spin estimates}

Figure~\ref{fig:corner} shows the two-dimensional posterior distributions for the remnant mass and spin of GW150914 under different ringdown models, together with those obtained from the full IMR analysis. Color contours denote the 90\% credible regions obtained for different nine values of the offset time, $\tstartgeo = t^\mathrm{geo}_\mathrm{start} - t^\mathrm{geo}_\mathrm{peak}$, measured in units of the remnant black-hole mass from the best-fit \ac{imr} waveform. The models compared are $\fundamental$, $\overtone$, $\mathrm{IMR}$, $\mathrm{IM\mbox{-}R}_{22}$, and $\mathrm{IM\mbox{-}R}$. While $\mathrm{IM\mbox{-}R}_{22}$ includes only the dominant quadrupole modes of \nrsur, both $\mathrm{IMR}$ and $\mathrm{IM\mbox{-}R}$ incorporate all modes with $\ell \leq 4$. Finally, black contours correspond to the values inferred from the full IMR analysis. In all cases, the IM-R waveforms are obtained as the portion of the full \texttt{NRSur7dq4} waveform starting at a time $t_{\rm start}^{\rm waveform} = \tstartgeo$ after the merger time, which is defined in units of remnant black hole mass. In future work, we will explore the impact of leaving $t_{\rm start}^{\rm waveform}$ as a freely sampled parameter.

Consistent with~\citet{Cotesta:2022pci}, we find that for all $\tstartgeo/M \in [-6,1]$, the $\overtone$ posteriors lie closer to the remnant properties inferred from the full \ac{imr} analysis than those obtained through $\fundamental$. This apparent agreement, however, is not necessarily physically meaningful through the whole range of starting times since, in principle, a common horizon may not yet have formed at times $\tstartgeo/M < -2$, for which the $\overtone$ estimates are inconsistent with those inferred from the IMR analysis. In contrast, for times $\tstartgeo/M \in [-2,4]$, one can see how the $\overtone$ model now yields results consistent with IMR ones, while the $\fundamental$ yields inconsistent ones, suggesting both the start of the linear regime and the actual presence of a detectable overtone.
For times $\tstartgeo/M \geq 4$, both the $\overtone$ and $\fundamental$ models yield comparable results, with the former showing significantly broader credible regions due to its larger freedom. As we will discuss next, this also leads to an important Ockham penalty that \textit{makes the detection of the overtone through this method not statistically significant at any of the considered starting times}.

In contrast to the above models, the IM-R approach leads to both significantly more stringent credible contours in agreement with the full IMR analysis and a significantly smaller Ockham penalty, which we will focus on in the next section. Both these aspects arise from the optimal prior for remnants of BBHs that this model effectively imposes on the QNM parameters.   

    \subsection{Testing the no-hair theorem using overtones and overtone recovery}

\begin{figure}
    \centering
    \includegraphics[width=0.49\textwidth]{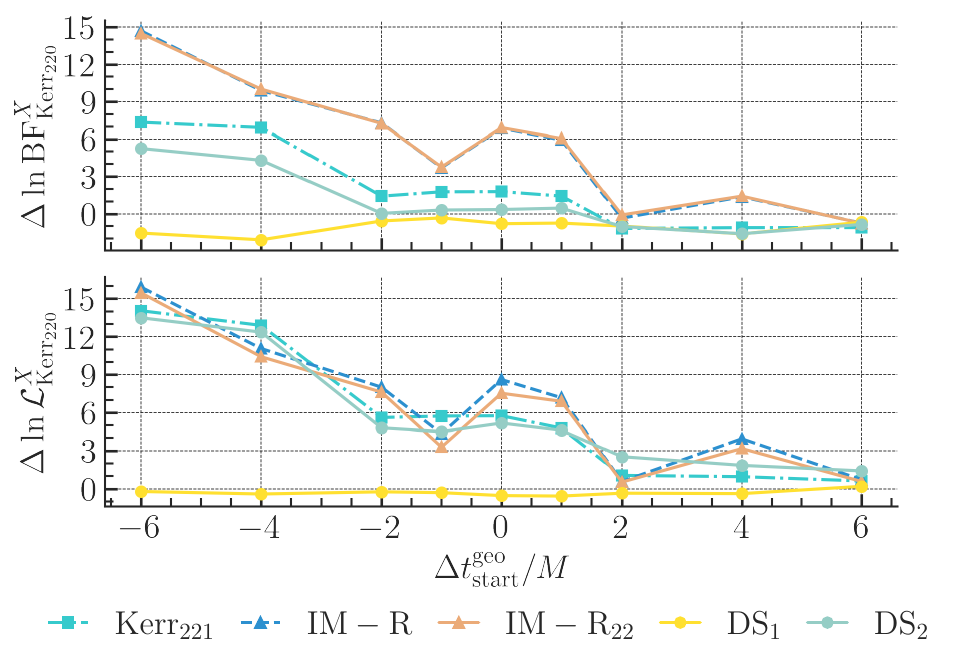} \\
    \caption{Log Bayes factor and maximum Log Likelihood for our different ringdown models with respect to the Kerr${}_{220}$ model as a function of the starting time of the analysis.}
    \label{fig:logb}
\end{figure}

Here we discuss the presence of the $(2,2,1)$ overtone in the GW150914 ringdown and the extent to which this overtone allows us to test the no-hair theorem.

The top panel of Figure~\ref{fig:logb} shows the difference of the natural logarithm of the Bayes Factor for our different ringdown models vs. the baseline $\fundamental$ model as a function of offset times, $\tstartgeo = t^\mathrm{geo}_\mathrm{start} - t^\mathrm{geo}_\mathrm{peak}$, expressed in units of the final \ac{bh} mass. It is clear that for starting times larger than $-2M$ the maximum evidence we find for the presence of the $(2,2,1)$ overtone is around $ \ln \mathrm{BF}^{\overtone}_{\fundamental} \simeq 1.4$ or, in linear terms, $\mathrm{BF}^{\overtone}_{\mathrm{Kerr}{220}} \simeq 4$. This is, we find only very mild evidence for such an overtone. The bottom panel of Figure~\ref{fig:logb}, however, shows that such mild evidence is due to the larger parameter space covered by the \( \overtone \) model. In particular, the panel shows that for starting times between $-2M$ and $1M$ the maximum log-likelihood obtained by the \(\overtone\) model is always five units larger than that attained by the \(\fundamental\) model. This means that the corresponding best-fitting ${\overtone}$ template is nearly 150 times more supported by the data than the corresponding ${\fundamental}$ template. However, this gain in likelihood is suppressed by the large Ockham penalty paid by the ${\overtone}$ model.

Next, we test the no-hair theorem using the Kerr ringdown models by computing the log Bayes Factor with respect to \ac{qnm} models consisting of different numbers of free damped sinusoids, which do not comply with the no-hair theorem. First, the two-damped sinusoid model (\(\DStwo\)) always reaches a maximum likelihood consistent with that of the $\overtone$ model. This is expected, as the latter model is contained in the former. However, the larger freedom of the \(\DStwo\) model results in a slightly higher Ockham penalty, leading to a lower Bayes Factor. Such a difference, however, never surpasses a very mild value of $\ln \mathrm{BF}^{\overtone}_{\DStwo} \simeq 1$, or $ \mathrm{BF}^{\overtone}_{\DStwo} \simeq 2.7$. Moreover, even if restricting the DS model to a single damped sinusoid, we only obtain $ \mathrm{BF}^{\overtone}_{\DSone} \simeq 7$. In other words, \textit{a semi-agnostic ringdown analysis can barely distinguish between the scenario of a black-hole radiating through two \acp{qnm} and some potential exotic compact objects \citep{Cardoso2019_livingreview, NicoMiguel_review} that do not comply with the no-hair theorem radiating through either two or one \acp{qnm}.}

These findings highlight a key limitation of (semi-)agnostic \ac{qnm} recovery: an increasing number of \acp{qnm} requires an increasing number of model parameters, which increases the Ockham penalty and prevents a strong preference for such a model. In contrast, \ac{qnm} models with amplitudes and phases informed by \ac{bbh} physics avoid this penalty, as shown next.

\subsection{Testing the no-hair theorem using the IM-R model}
Next, we present the results obtained with the IM-R ringdown model. This model effectively incorporates all possible \acp{qnm} with $l \leq 4$, while having the corresponding amplitudes and phases restricted to those consistent with remnants of quasi-spherical \acp{bbh}. If the remnant is indeed a black hole formed in such a merger, the IM-R model should provide the optimal description of the data, maximizing the likelihood using a minimal parameter space.

Figure~\ref{fig:logb} supports this expectation. The top panel shows that IM-R yields the largest log-likelihood, exceeding that of the Kerr and DS models including two \acp{qnm} by $\sim 3$ units; and surpassing single-mode models by \(\sim 9\) units for starting times between $0M$ and $1M$. The bottom panel further demonstrates that the Ockham penalty does not remove this likelihood difference: the IM-R model achieves $\ln \mathrm{BF}^{\text{IM-R}}_{\fundamental} \simeq 7$, or $\mathrm{BF}^{\text{IM-R}}_{\fundamental} \simeq 10^3$. Thus, the IM-R model is \textit{decisively preferred} over the $\fundamental$ model.

Following~\citet{CalderonBustillo:2020rmh}, we test the no-hair theorem by comparing IM-R with the most favored DS model, finding $\mathrm{BF}^{\text{IM-R}}_{\DStwo} \gtrsim 650$ (a $0.15\%$ violation probability), thereby confirming their result with a more robust data treatment.

\subsection{No evidence for higher-order angular modes}

Since \nrsur does not provide an explicit \ac{qnm} decomposition of its ringdown, we cannot directly use it to test the presence of overtones in the data. However, this model can be decomposed into angular modes $h_{\ell m}(t)$, which allows us to study the presence of \ac{gw} modes beyond the dominant quadrupole $h_{2,\pm 2}$ in the post-merger portion of the data. To this end, we compare the data with the IM-R model and its quadrupole-only variant IM-R$_{22}$. As shown in Figure~\ref{fig:logb}, the full model achieves a slightly higher maximum log-likelihood at $\tstartgeo=0M$, but both yield identical Bayes factors at all times, indicating no evidence for angular modes beyond the quadrupole.

\subsection{Validity of the 1-overtone model and parameter recovery}

Here, we assess the validity of the $\fundamental$ and $\overtone$ models by examining remnant mass and spin estimates at different analysis start times, followed by the consistency of inferred mode amplitudes.

  \subsubsection{Remnant parameters} 

\begin{figure}
    \centering
    \includegraphics[width=0.5\textwidth]{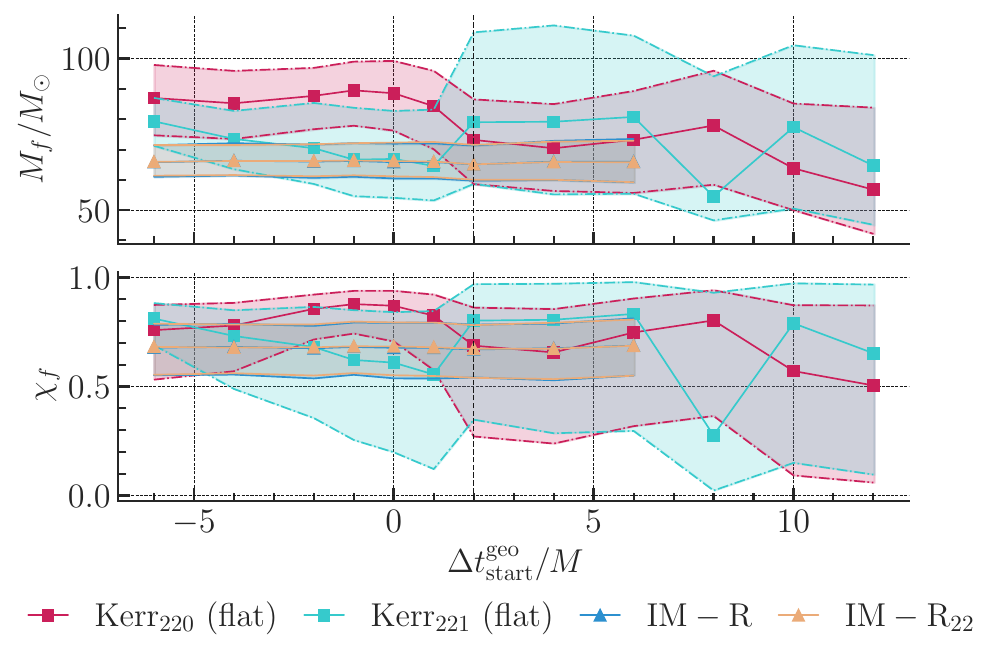}
    \caption{\textbf{Time evolution of final mass and spin estimates}. We show the 1-dimensional $90\%$ credible intervals for the final mass and spin of GW150914 for several choices of the starting time of the analysis, according to different ringdown models. The vertical line denotes the time $2M$ at which the evidence for all models becomes comparable. In particular, any evidence for the $(2,2,1)$ overtone disappears after that.}
\label{fig:mf_evol_1d}
\end{figure}

Figure~\ref{fig:mf_evol_1d} shows the one-dimensional $90\%$ credible intervals for the final mass and spin of GW150914 inferred with the IM-R and Kerr models for different starting times of the analysis. The vertical dashed line at $\tstartgeo=2M$ marks the point where all models yield a comparable Bayesian evidence. Beyond this time, there is no support for the $\overtone$ model, and the log-likelihood differences between all models are negligible. 

As expected, IM-R results remain consistent across different start times, with uncertainties growing as less data is included. For early times ($-2.5M$ to $1M$), estimates from the fundamental-only model are inconsistent with IM-R. In contrast, the $\overtone$ model yields consistent mass and spin posteriors, along with a modest evidence gain and improved fit to the data with respect to the ${\fundamental}$ model. 

Notably, the mentioned preference for the \(\overtone \) model disappears around $2M$, after which the \(\fundamental\) mode results are more visibly consistent with those of the IM-R model than the \(\overtone\) ones. This behaviour supports the expectation that any overtones associated with the $(2,2,0)$ mode have effectively decayed after $2M$.

\subsubsection{QNM amplitudes} 

\begin{figure*}
    \centering
    \includegraphics[width=0.98\textwidth]{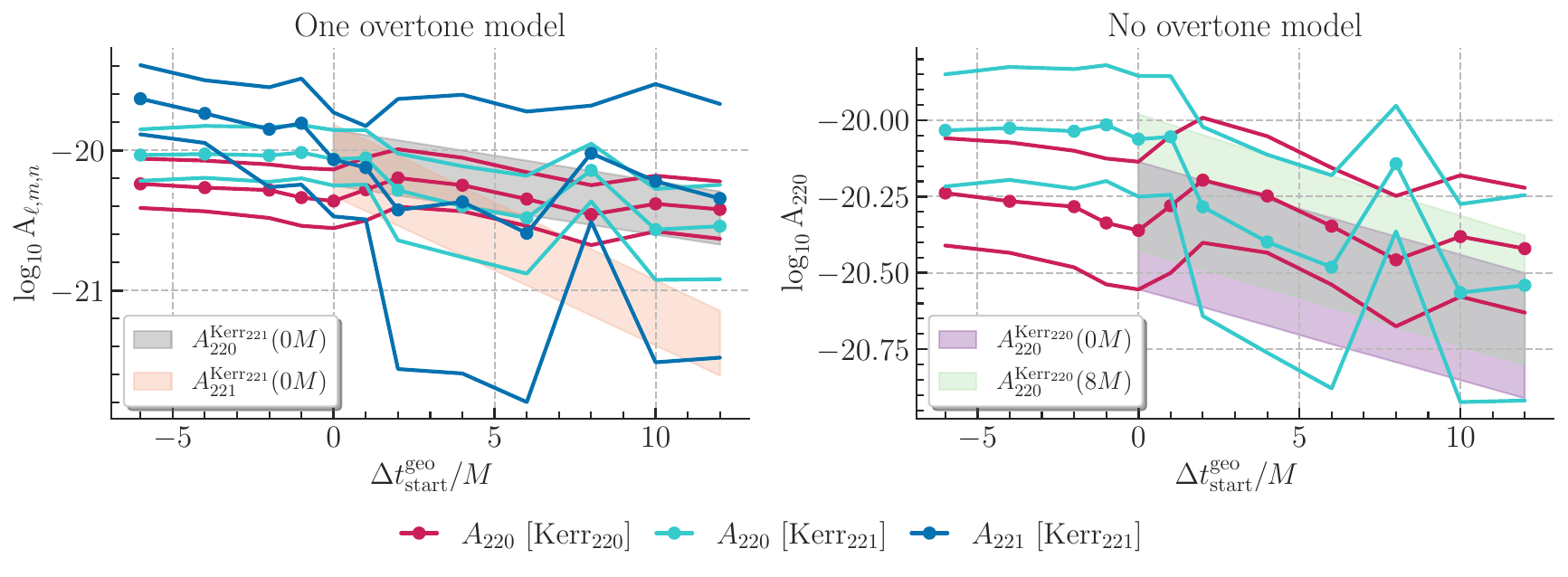}
    \caption{\textbf{Self-consistency of quasi-normal mode amplitudes estimations for GW150914}. The scatters and empty contours on the left panel denote the median and $68\%$ credible intervals for the amplitudes of the fundamental ringdown mode and its overtone estimated with either the \(\fundamental\) (left and right) or $\overtone$ (only left) models for different choices of the starting time of the analysis.  The shaded regions on the left panel denote the $68\%$ credible intervals for the fundamental mode (grey) and its overtone (orange) estimated using the amplitude and decay times measured at $\tstartgeo=0M$ using the $\overtone$ model. The right panel shows the corresponding estimates for the amplitude of the fundamental mode based on the amplitude and damping time measurements performed with the $\fundamental$ model at times $\tstartgeo=0M$ (purple) and $\tstartgeo=8M$ (green).}
\label{fig:amp_consistency}
\end{figure*}

We now examine the self-consistency of the mode amplitudes inferred by our two Kerr models across different $\tstartgeo$. Specifically, we compare these amplitudes with predictions based on amplitude and damping-time estimates of these models obtained at $\tstartgeo=0M$, when the $\overtone$ model is expected to describe the data better, and at $\tstartgeo=8M$, when the overtone should have turned off, making the $\fundamental$ model become valid.  

The left panel of Figure~\ref{fig:amp_consistency} shows the $68\%$ credible intervals for the base-10 logarithm of the fundamental $(2,2,0)$ mode amplitude inferred through the $\fundamental$ and $\overtone$ models, together with the overtone amplitude. The shaded bands denote the fundamental (grey) and overtone (orange) amplitudes predictions based on measurements at $\tstartgeo=0M$ performed with the $\overtone$ model. In the right panel, we show the same measurements of the fundamental mode amplitude as in the left panel, but we now compare these to theoretical predictions based on the estimates of the amplitude and damping times obtained through the $\fundamental$ model at times $\tstartgeo=0M$ and $\tstartgeo=8M$.

First, it is clear that the overtone amplitudes (blue) are reliably predicted for all times. In particular, median estimated values are well centered within theoretical predictions up to $\tstartgeo=2M$ -- which is the latest time in Figure~\ref{fig:logb} up to which the $\overtone$ model is marginally favoured -- and stay within such predictions until $\tstartgeo=6M$. After that time, while median estimated values fall outside theoretical predictions, the $68\%$ becomes so large due to the overtone weakness that estimates can be considered consistent. Second, the amplitude of the fundamental mode predicted by the $\overtone$ model based on estimates at $\tstartgeo=0M$ is consistent with those estimated through the same $\overtone$ model at all times and with those estimated through the $\fundamental$ model for times $\tstartgeo \geq 2M$ when, again, any evidence for the presence of the overtone is lost. In conclusion, these results show that overtone and fundamental-mode amplitude estimates performed at $\tstartgeo=0$, when the overtone is conjectured to be active, seem to be self-consistent. 

By contrast, the right panel shows that the fundamental mode amplitudes predicted by the \(\fundamental\) model at \(\tstartgeo=0M\) (purple) barely match those measured at times  $\tstartgeo \geq 2M$ using the same model (red), indicating inconsistency. In contrast, the amplitudes predicted for the fundamental mode based on estimates performed at $\tstartgeo=8M$(green), when the overtone is expected to have vanished, are entirely consistent with those measured by both ringdown models in the range $0- 12M$.  

As a summary, we find that the $\overtone$ model at early times and the $\fundamental$ model at late times both give self-consistent and physically meaningful amplitude estimates. By contrast, applying the $\fundamental$ model too early leads to inconsistent fundamental amplitudes, confirming that an overtone is needed there or possibly reflecting non-linearities in the post-peak signal immediately after merger.

\begin{figure*}
    \centering
    \includegraphics[width=0.98\textwidth]{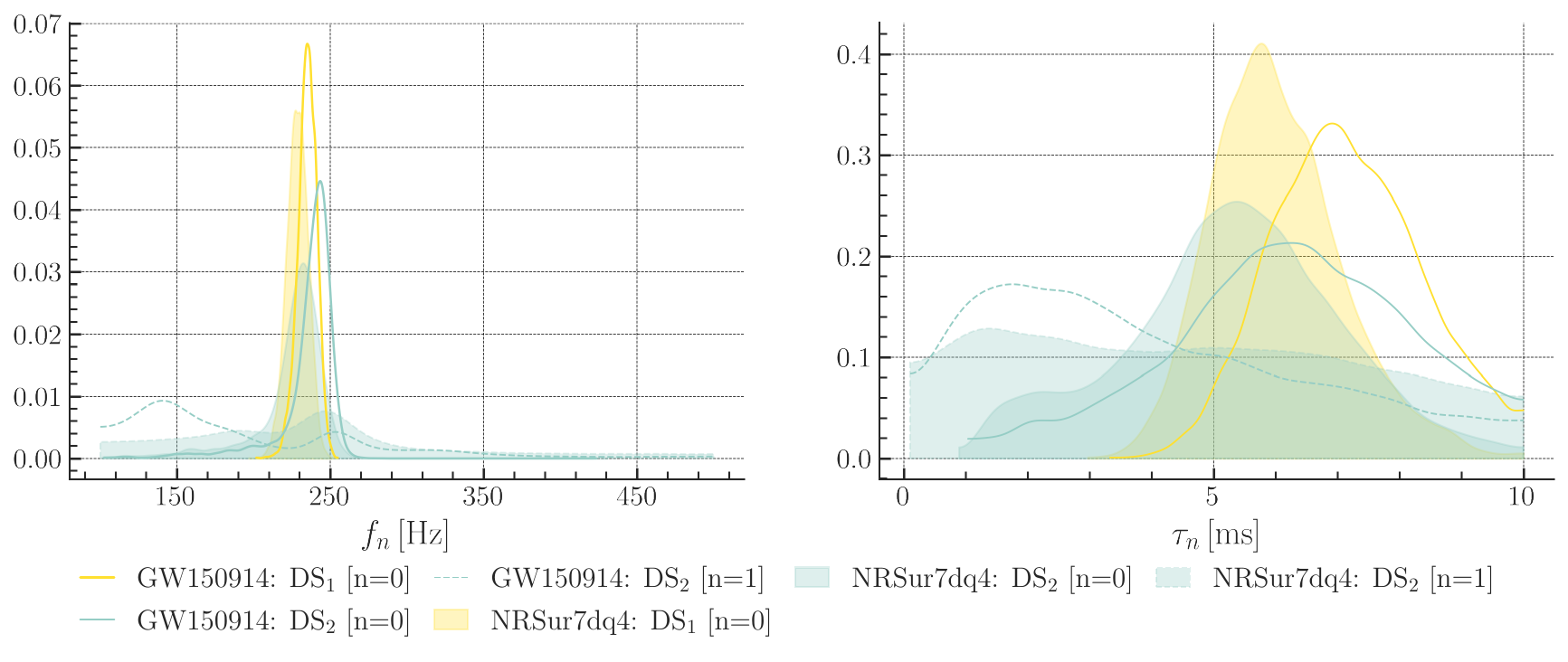}
    \caption{\textbf{Frequency and damping times using an agnostic ringdown model on GW510914 and a consistent injection}. We show the posterior distributions for the mode frequencies (left) and damping times (right) inferred under our 1 and 2 Damped-sinusoid models for the case of GW150914 (empty bounded KDE) and a simulated signal consistent with it (filled bounded KDE). The starting time of the analysis is $\tstartgeo = 0M$.}
\label{fig:freq_taus}
\end{figure*}

\subsection{Agnostic recovery of the GW150914 ringdown}

Figure~\ref{fig:freq_taus} shows the posterior distributions for the central frequencies and damping times inferred with one- (\(\DSone\), orange) and two-mode (\(\DStwo\), navy) damped-sinusoid models at $\tstartgeo=0M$. Unfilled distributions correspond to GW150914 data, while filled distributions correspond to the analysis of a simulated \nrsur injection with consistent parameters injected in zero noise. Both analyses yield consistent results.

Under the $\DSone$ model, we obtain $(f_0,\tau_0) = (229^{+11}_{-12}\,\mathrm{Hz},\,5.9^{+2.0}_{-1.5}\,\mathrm{ms})$. This differs significantly from the LVK estimate $(251^{+8}_{-8}\,\mathrm{Hz},\,4.0^{+0.3}_{-0.3}\,\mathrm{ms})$~\cite{LIGOScientific:2016lio}. This is, however, expected: The LVK was performed at a time when a single \acp{qnm} is expected to be active, whereas our analysis started much earlier, when the signal is expected to have contributions from more \acp{qnm}.

Therefore, we consider the two-damped sinusoid $\DStwo$ to model the post-merger signal. This yields $(f_0,\tau_0) = (230^{+19}_{-59}\,\mathrm{Hz},\,5.3^{+2.7}_{-3.2}\,\mathrm{ms})$ and $(f_1,\tau_1) = (237^{+196}_{-118}\,\mathrm{Hz},\,4.5^{+4.7}_{-3.9}\,\mathrm{ms})$. At face value, these appear inconsistent with the results of~\citet{CalderonBustillo:2020rmh}, who report $(f_0,\tau_0) = (254^{+21}_{-9}\,\mathrm{Hz},\sim5\,\mathrm{ms})$ and $(f_1,\tau_1) = (168^{+28}_{-19}\,\mathrm{Hz},\sim4\,\mathrm{ms})$. Here, \(f_0\) matches the expected central frequency of the fundamental mode of GW150914, and \(f_1\) roughly matches GW150914's merger frequency.

Such apparent discrepancy, however, arises from different labelling conventions: in this work, the ``0'' mode is defined as the largest-amplitude one, whereas ~\citet{CalderonBustillo:2020rmh} assigned ``0'' to the higher-frequency component and further constrains it to have a longer damping time, thereby reducing the parameter space their model covers. 

Relabelling our modes accordingly, we find $(f^{*}_0,\tau^{*}_0) = (246^{+106}_{-13}\,\mathrm{Hz},\,6.4^{+3.0}_{-5.3}\,\mathrm{ms})$ and $(f^{*}_1,\tau^{*}_1) = (164^{+74}_{-54}\,\mathrm{Hz},\,3.7^{+4.4}_{-2.7}\,\mathrm{ms})$, consistent with~\citet{CalderonBustillo:2020rmh}. The same procedure applied to the \(\overtone\) analysis of the simulated \nrsur signal which yields $(f^{*}_0,\tau^{*}_0) = (246^{+187}_{-21}\,\mathrm{Hz},\,5.7^{+3.4}_{-4.9}\,\mathrm{ms})$ and $(f^{*}_1,\tau^{*}_1) = (205^{+31}_{-89}\,\mathrm{Hz},\,4.5^{+3.7}_{-3.3}\,\mathrm{ms})$. This level of agreement between real and simulated data reinforces the robustness of our analysis and highlights the importance of careful mode labelling and model assumptions when interpreting ringdown spectra.

\subsection{Physical interpretation: non-orthogonality of QNMs and feasibility of agnostic black-hole spectroscopy}

\begin{figure}
    \centering        
    \includegraphics[width=0.49\textwidth]{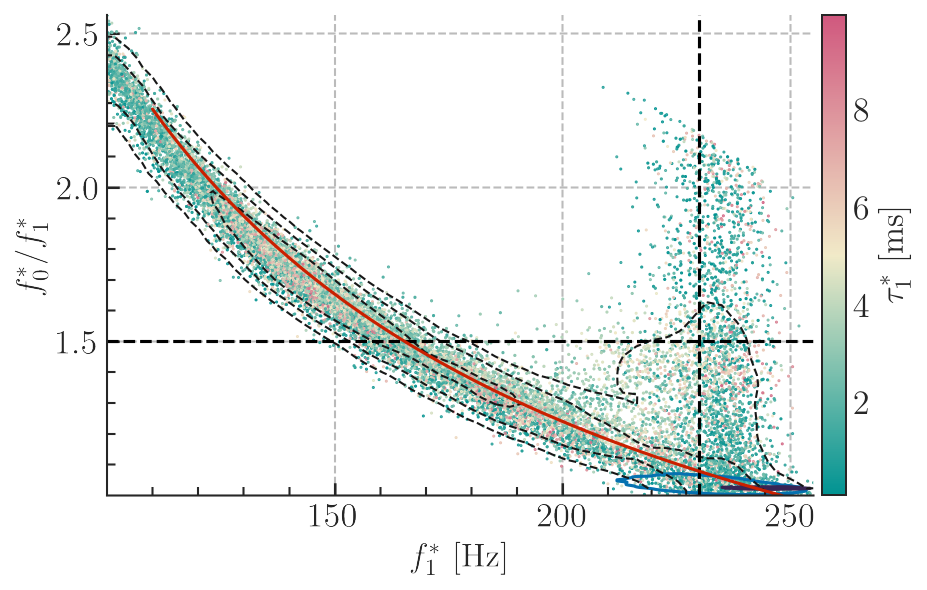}\\
    \includegraphics[width=0.49\textwidth]{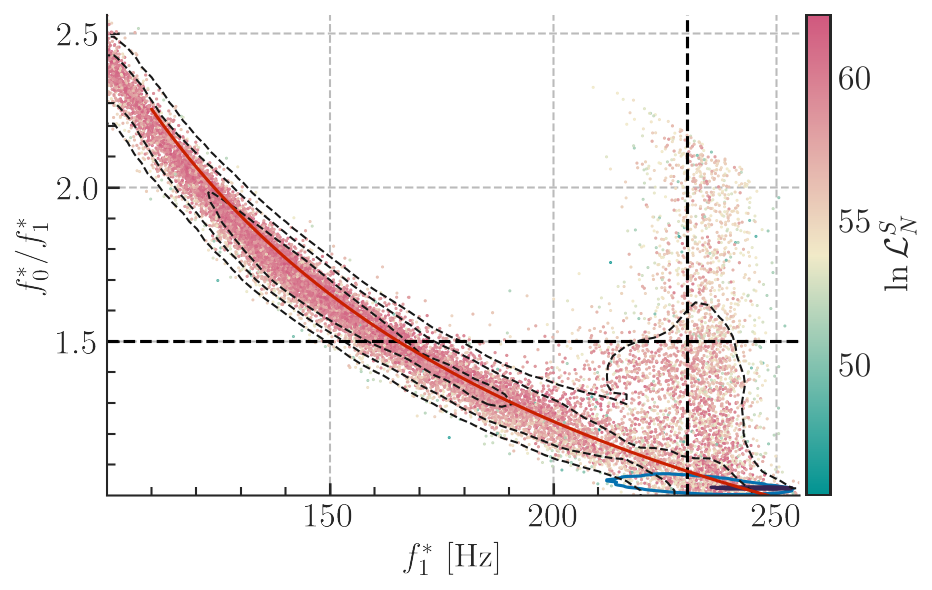}
    \caption{Posterior samples from the agnostic damped-sinusoid analysis of GW150914 in the $(f_1^\ast, f_0^\ast/f_1^\ast)$ plane. Black dashed contours mark the $30\%$, $68\%$, and $90\%$ credible regions. \textit{Top}: points colored by the shorter damping time $\tau_1^\ast$. \textit{Bottom}: points colored by $\ln \mathcal{L}^S_N$. The vertical dashed line indicates $f_1^\ast\simeq246\,\mathrm{Hz}$, consistent with the fundamental $(2,2,0)$ mode, while the horizontal dashed line marks $f_0^\ast/f_1^\ast=1.5$, expected for the $(3,3,0)$ mode. The results highlight three possible interpretations of the ringdown: fundamental plus overtone, fundamental plus merger frequency, and fundamental plus higher-order mode. The blue and solid black contour (at the right bottom corner) encloses the corresponding predictions from the \(\overtone\) and IM-R model.}
\label{fig:nested_plot-1}
\end{figure}

\begin{figure}
    \centering        
    \includegraphics[width=0.49\textwidth]{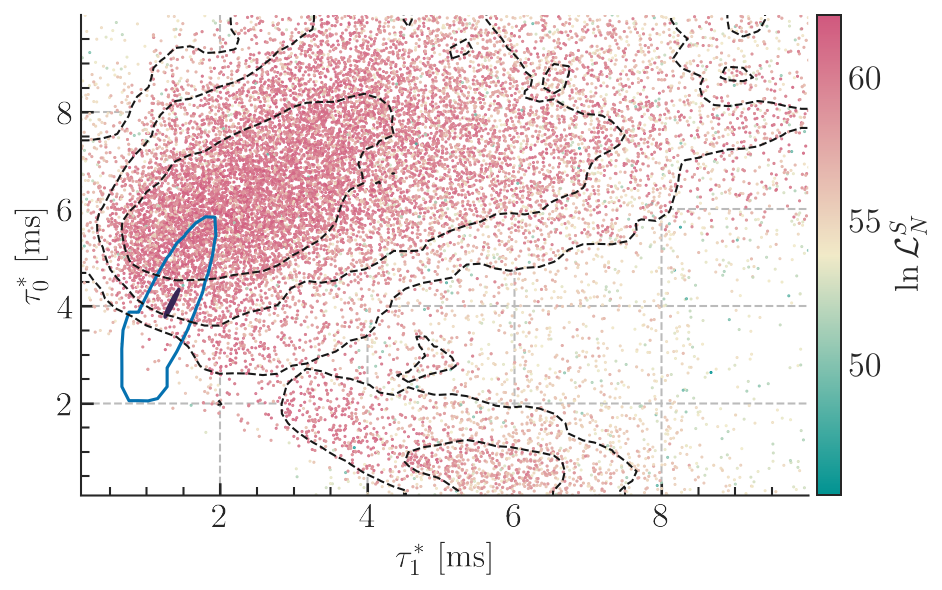}
    \caption{Posterior samples from the agnostic damped-sinusoid analysis of GW150914 in the $(\tau_{1}^\ast, \tau_{0}^\ast)$ plane, where $\tau_{1}^\ast$ and $\tau_{0}^\ast$ are the shorter- and longer-lived damping times, respectively. Black dashed contours denote the $30\%$, $68\%$, and $90\%$ credible regions. Points are colored by log-likelihood, illustrating that comparable likelihoods are obtained both within and outside the regions consistent with the no-hair theorem. This reflects the flexibility of the agnostic model and the non-uniqueness of individual mode recovery. Here also, the blue and solid black contour encloses the corresponding predictions from the \(\overtone\) and IM-R model.}
\label{fig:tau_plot}
\end{figure}

We now try to provide a physical interpretation of the frequencies and damping times recovered using the agnostic damped-sinusoid models and compare them with those obtained from the Kerr and IM-R analyses. We recall that the latter two models may be interpreted as increasingly constrained realizations of the agnostic approach: Kerr enforces \ac{qnm} frequencies and damping times to be determined by the final mass and spin, while IM-R further determines mode amplitudes and phases through the progenitor \ac{bbh} parameters. Therefore, despite their different parametrizations, the three models are nested, differing primarily in the freedom they allow for fitting the data.  

The two panels of Figure~\ref{fig:nested_plot-1} shows posterior samples obtained at \(\tstartgeo=0M\) for the frequency of the lowest-frequency mode $f^{*}_{1}$ and the ratio $f^{*}_{0}/f^{*}_{1}$, where $f^{*}_{0}$ denotes the frequency of the higher-frequency mode. The color scale indicates either the shorter damping time $\tau_{1}^\ast$ (top) or the log-likelihood $\ln \mathcal{L}^S_N$ (bottom). The red line marks $f^{*}_{0}=246\,\mathrm{Hz}$, which corresponds to the median value inferred from the damped-sinusoid analysis. Blue and solid black contours indicate the $(2,2,0)$ and $(2,2,1)$ mode frequencies predicted from the final mass and spin measured with the IM-R and $\overtone$ models, using the fitting relations in~\citet{Berti:2006wq}.

Next, Figure~\ref{fig:tau_plot} shows the posterior distribution of the shorter- and longer-lived damping times $(\tau_{1}^\ast, \tau_{0}^\ast)$ obtained with the agnostic damped-sinusoid model. Similar to the previous plot, the contours mark the $30\%$, $68\%$, and $90\%$ credible regions, while points are colored by log-likelihood. We find that regions consistent with the no-hair theorem (blue contour) achieve likelihoods comparable to those outside it, highlighting both the flexibility of the agnostic model and the non-uniqueness of individual mode recovery.

Both figures clearly demonstrate that the three models lead to vastly different combinations of frequencies and damping times because they impose significantly different constraints (effectively, different priors) on these parameters. A closer look at Figure~\ref{fig:nested_plot-1} reveals that the agnostic model roughly admits three distinct ways to reproduce the ringdown of GW150914. First, in the bottom-right region, the data is fitted by a combination of a ringdown mode with frequency $\sim 250\rm Hz$ -- expected for the fundamental mode -- and a second mode of similar frequency and small damping time $\sim 2 \rm ms$, as expected for its overtone. Unsurprisingly, this matches well with the predictions from the $\overtone$ and IM-R analyses.
Second, along the diagonal region on the mid/left side of the plot, the data is fitted through a combination of a short-lived mode with a frequency $\sim (130-180)\rm Hz$, roughly consistent with the merger frequency of GW150914, combined with a longer-lived mode with a frequency $\sim 250 \rm Hz$, again consistent with expectations for the fundamental ringdown mode. 
\textit{Although incompatible with the no-hair theorem}, this outcome is physically understandable: the model recovers the longest-lived mode while making the second match the merger frequency, which coincides with the peak of the signal amplitude. One physical interpretation is that GW150914 may contain an infinite set of unresolved \acp{qnm}. In this situation, the damped-sinusoid model effectively captures such a mode combination with one of its modes. This is consistent with the larger maximum likelihood reached by IM-R, which incorporates the full \ac{qnm} spectrum. A third branch, in the mid-right region, admits a more intriguing interpretation: two long-lived modes, one with a `biased' fundamental frequency $f^{*}_{1}\sim 230\,\mathrm{Hz}$ and another at $f^{*}_{0}\sim 1.5 f^{*}_{1}$, somewhat consistent with expectations for the $(3,3,0)$ mode.

\begin{figure}
    \centering
    \includegraphics[width=0.48\textwidth]{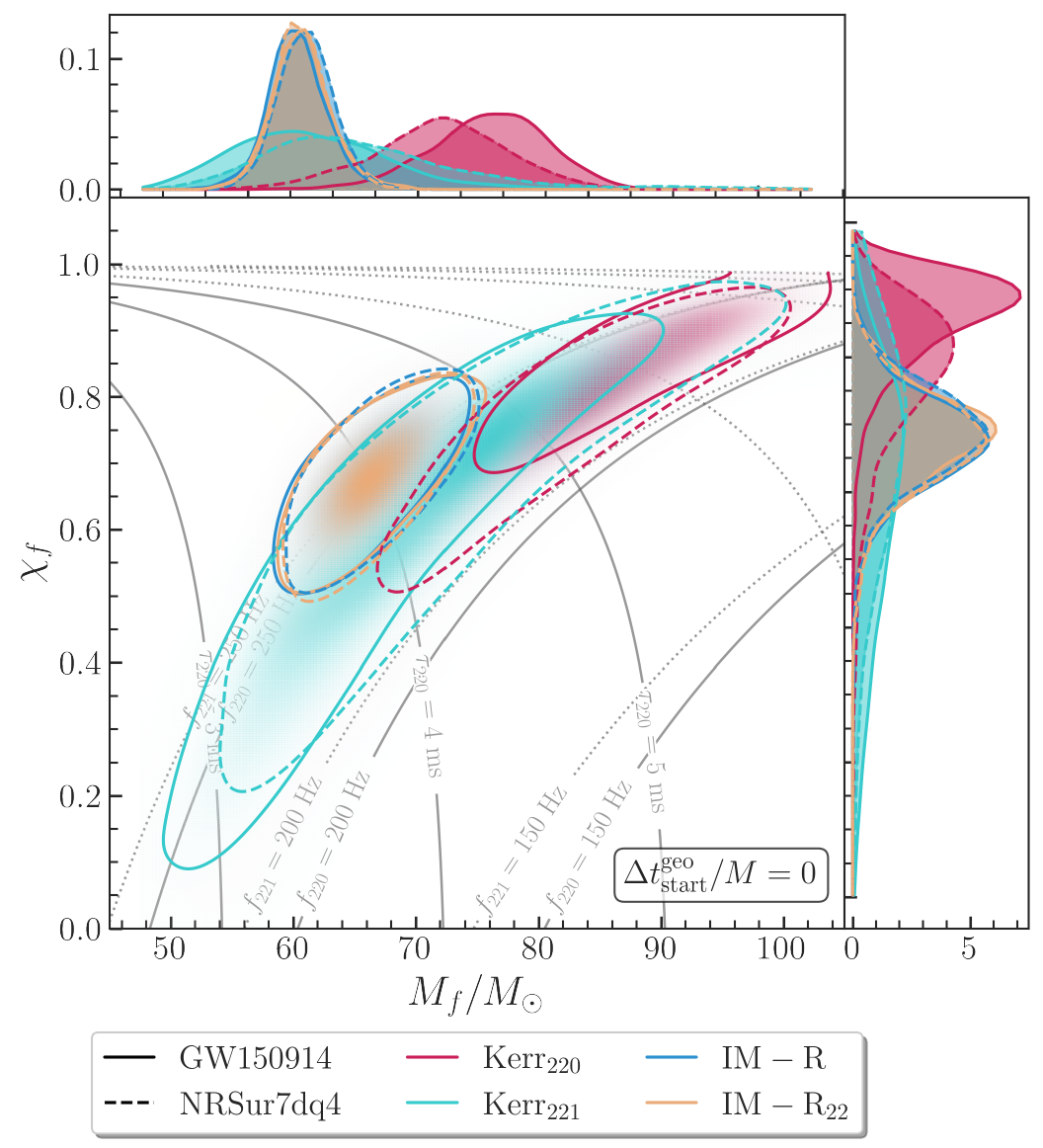}\\
    \includegraphics[width=0.48\textwidth]{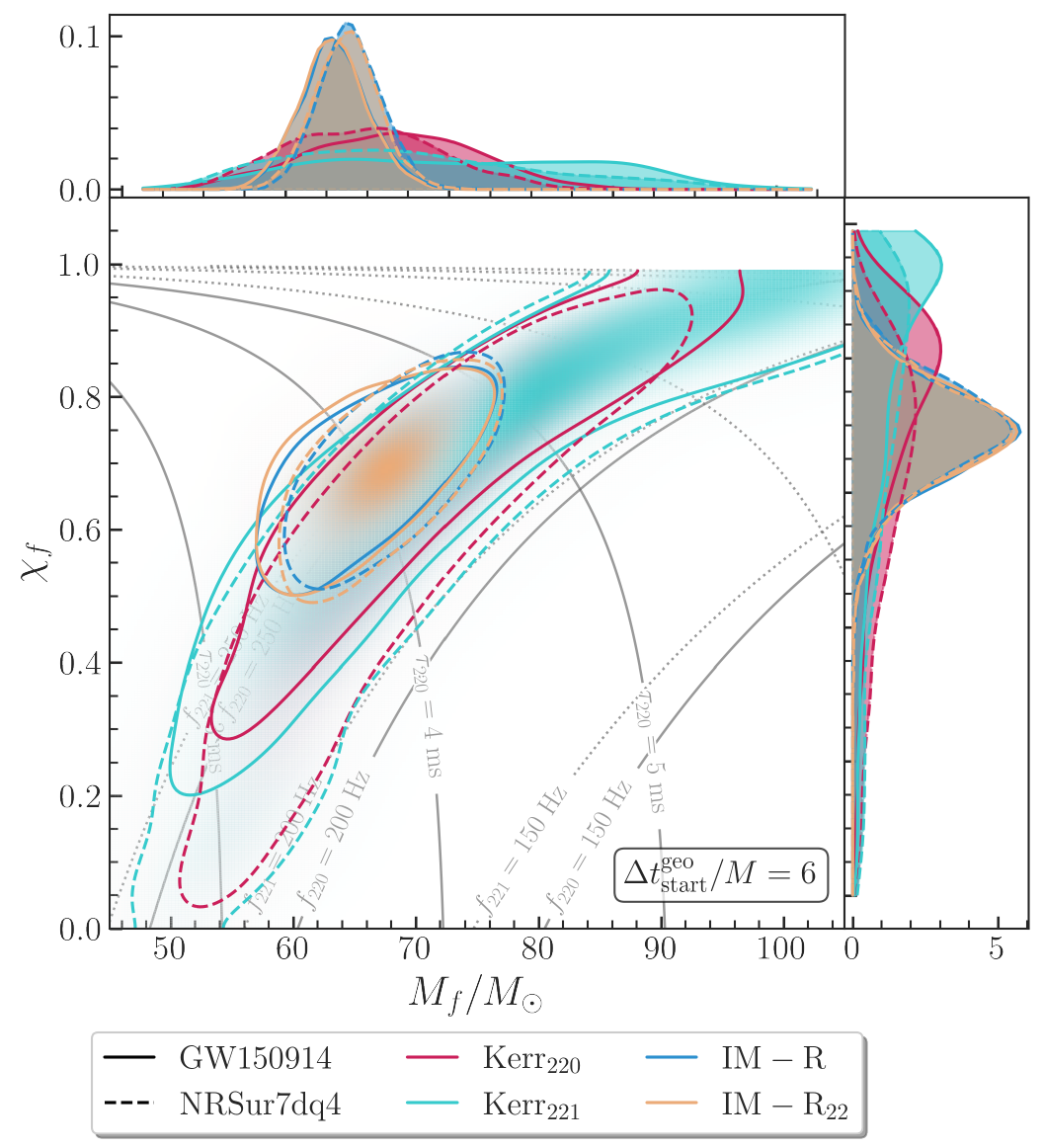}
    
    \caption{Comparison of the remnant mass $M_f$ and spin $\chi_f$ inferred from GW150914 (real data) and an \nrsur injection (simulated data), using different template models. Results are shown for ringdown starting times $\Delta t_{\mathrm{start}}^{\mathrm{geo}}=0M$ (top) and $\Delta t_{\mathrm{start}}^{\mathrm{geo}}=6M$ (bottom).}
    \label{fig:compare-results}
\end{figure}

In summary, our agnostic analysis indicates that there are, at least, three ways to reproduce the ringdown of GW150914 via damped sinusoids. One branch where the modes reproduce the ``true'' fundamental mode and its overtone, one branch where one mode reproduces the ``true'' fundamental mode and the other tries to fit the merger frequency, and a third branch where the modes reproduce a ``biased'' fundamental mode and, roughly, the corresponding $(3,3,0)$ mode. The fact that the first two options clearly achieve similar likelihoods (in fact, the second one reaches a slightly larger one) demonstrates that the non-orthogonality of \acp{qnm} allows for describing a given \ac{qnm} combination with many possible ones. In our view, this is an explicit demonstration that individual mode recovery, which is the main ingredient of ``classical'' black-hole spectroscopy, is ill-defined, as the inferred \ac{qnm} content of a signal, by definition, depends on the \ac{qnm} model used. 

\section{Cross-validation of our results with a GW150914-like injection}
\label{sec:gw150914-like}

In this section, we assess the robustness of our results in the previous section by analyzing a simulated signal consistent with GW150914 generated with the \nrsur model injected in zero noise. 

\subsection{Parameter recovery}

Figure~\ref{fig:compare-results} compares the remnant mass $M_f$ and spin $\chi_f$ inferred from GW150914 (solid) and our \nrsur injection (dashed) under different models. We show this for two analysis start times $\tstartgeo/M=0$ (top) and $\tstartgeo/M=6$ (bottom). The two sets of posteriors are in excellent agreement across models, confirming the robustness of our analysis. We note that similar consistency is also observed for other start times. 

\begin{figure}
    \centering
    \includegraphics[width=0.49\textwidth]{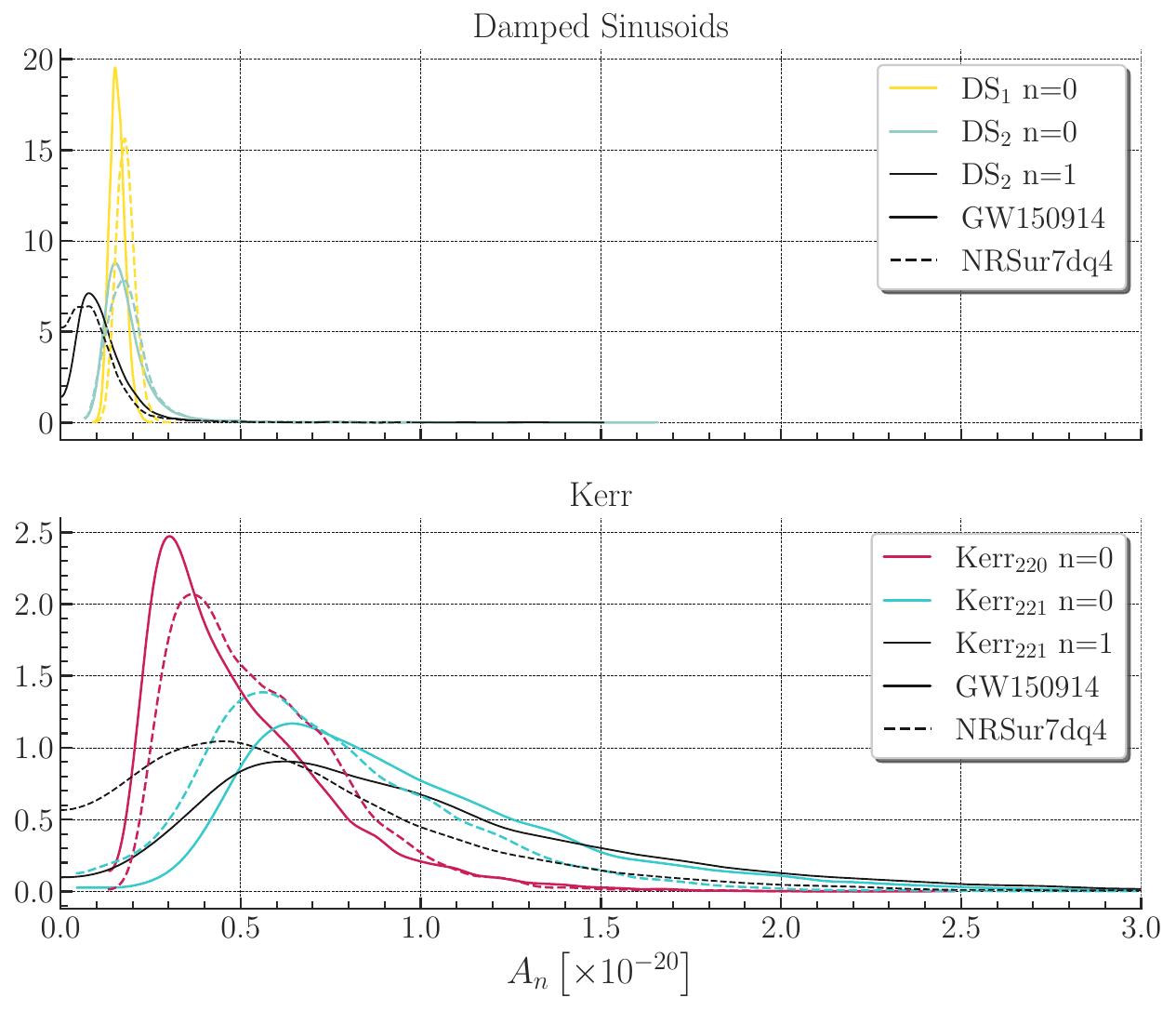} \\
    \caption{Posterior distributions of mode amplitudes $A_n$ inferred from GW150914 (solid) and from an NRSur7dq4 injection mimicking the event (dashed). \textit{Top}: results using damped-sinusoid (DS) models. \textit{Bottom}: results using Kerr-based models. In all cases, the amplitude posteriors inferred from the real signal closely match those from the synthetic injection, demonstrating the consistency of our analysis and the robustness of mode recovery across models.}

    \label{fig:Amps_comparison}
\end{figure}

We have further verified that the mode amplitude posteriors obtained under the Kerr and damped-sinusoid models agree for both the real and simulated signals (Figure~\ref{fig:Amps_comparison}), and that the recovered frequencies and damping times are consistent across analyses (Figure~\ref{fig:freq_taus}). Together, these results demonstrate that our methodology recovers self-consistent remnant properties from both GW150914 and its simulated analog.  

\subsection{Evidence for overtones and higher-order modes}

\begin{figure}
    \centering
    \includegraphics[width=0.48\textwidth]{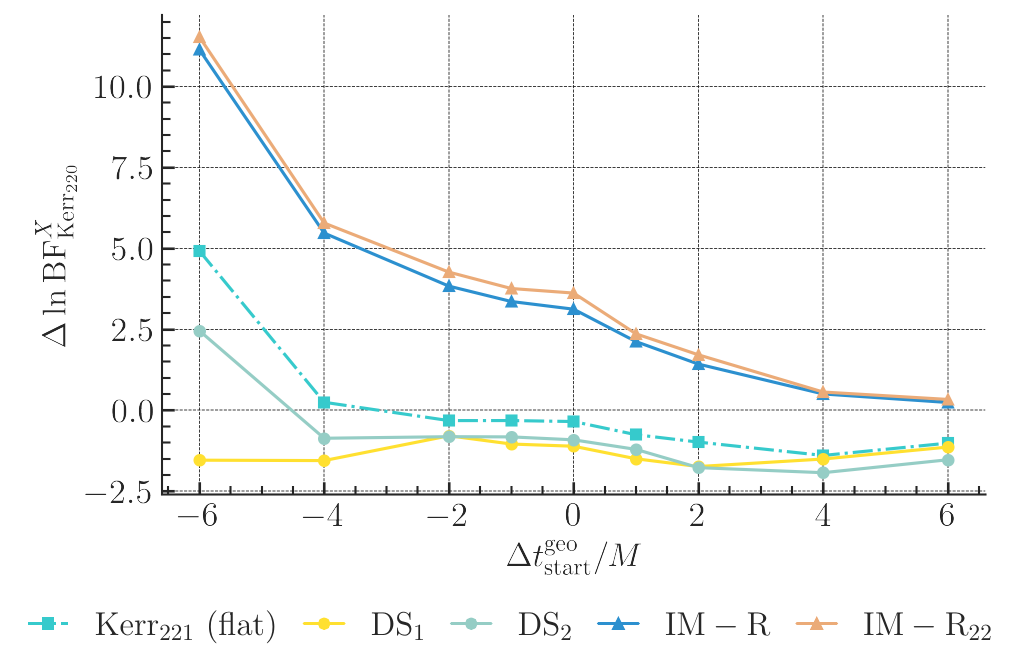}
    \caption{Log Bayes factor for a GW150914-like signal for different ringdown models with respect to the \(\fundamental\) model as a function of analysis start time.}
\label{fig:logb_injection_GW150914}
\end{figure}

We next examine the evidence for overtones and higher-order modes. Figure~\ref{fig:logb_injection_GW150914} shows the log Bayes factor obtained for our simulated signal, analogous to the bottom panel of Figure~\ref{fig:logb}. For \(\tstartgeo=0\), we find that the presence of an overtone is disfavoured with  $\ln \mathrm{BF}^{221}_{220} = -0.35$. This is consistent with the results reported in \citet{CalderonBustillo:2020rmh}, where the authors performed an analysis of a numerically simulated signal consistent with GW150914, as well as with our results on the actual GW150914 signal, which only yield weak evidence. Similarly, comparing the IM-R and IM-R$_{22}$ models yields $\ln \mathrm{BF}^{\mathrm{HM}}_{22} = -0.5$, disfavoring contributions beyond the quadrupole, in agreement with our GW150914 analysis.  

\subsection{Testing the no-hair theorem}

We now test the no-hair theorem using our synthetic signal. Comparing the preferred Kerr model ($\fundamental$) with the best-fitting damped-sinusoid model (\(\DSone\)) yields only mild support for the no-hair theorem, $\ln \mathrm{BF}^{\mathrm{No\text{-}Hair}}_{\mathrm{Hair}} = 1.11$. Adopting instead the IM-R model as the reference no-hair–compliant model strengthens the evidence to $\ln \mathrm{BF}^{\mathrm{No\text{-}Hair}}_{\mathrm{Hair}} = 3.7$, corresponding to odds of $\sim 40:1$ in favor of the no-hair theorem and nearly sub-percent violation probability of $2.4\%$, when assuming equal prior odds. These results are consistent with those obtained for GW150914 and with previous analyses of numerical simulations~\citet{CalderonBustillo:2020rmh}.

\section{Forecasting overtone-based spectroscopy using a large SNR injection}
\label{sec:high-snr}

As a final check, here we investigate the prospects for testing the no-hair theorem using overtones and semi-agnostic spectroscopy in the high-\ac{snr} regime. Specifically, we perform a zero-noise analysis of the numerically simulated signal consistent with GW150914. In particular, we choose the SXS simulation  \texttt{SXS:BBH:0305} which has aligned spins and therefore lacks any precession effects. We choose a sky-location, orientation, and a total mass consistent with the best-fit GW150914 signal reported in~\citet{Islam:2023zzj} but re-scale the source distance to yield an optimal post-merger network \ac{snr} of $\sim 100$, way beyond that of any ringdown signal observed to date. While somewhat unrealistic, the choice of an aligned-spin signal restricted to its quadrupole mode is motivated by the fact that this provides a controlled scenario to test the feasibility of no-hair theorem tests based on overtones of the fundamental $(2,2,0)$ mode. On the one hand, the restriction to the quadrupole mode prevents the presence of $(\ell,m) \neq (2,\pm 2)$ modes in the data. On the other hand, the lack of precession prevents asymmetries between the $(2,2)$ and $(2,-2)$ modes~\citep{Kamaretsos:2011um, Leong:2025raf} and mode-mixing~\citep{Bernuzzi:2010ty, Pan:2013rra, Berti:2014fga}, which would further complicate the analysis.

\subsection{Model selection and no-hair theorem tests}

\begin{table}[h]
    \centering
    \setlength{\tabcolsep}{8pt}
    \begin{tabular}{|c |c|c|}
        \hline
        Model & $\ln \mathrm{BF}^S_N$ & $\max(\ln \mathcal{L}^S_N)$ \\
        \hline
        \(\DSone\) & 5020.045 & 5035.056 \\
        \(\DStwo\)  & 5160.032 & 5187.407 \\
        $\mathrm{DS}_3$   & 5160.494 & 5190.828 \\
        $\mathrm{DS}_4$   & 5159.037 & 5189.938 \\
        \(\fundamental\) & 5020.146 & 5038.513 \\
        $\overtone$ & 5153.571 & 5178.468 \\
        $\mathrm{Kerr}_{222}$ & 5160.351 & 5191.081 \\
         $\mathrm{Kerr}_{223}$ & 5159.023 & 5190.897 \\
        IM-R$_{22}$ & {5167.396} & {5189.636} \\
        \hline
    \end{tabular}
    \caption{Log Bayes factors, $\ln \mathrm{BF}^S_N$ and maximum log-likelihood values, $\max(\ln \mathcal{L})$ at \(\tstartgeo=0M\) for different ringdown models, obtained from a GW150914-like \nrsur injection in zero noise with a post-merger SNR of 100.}
    \label{tab:SNR100}
\end{table}

Table~\ref{tab:SNR100} summarizes the model-selection results, reporting the Bayesian evidence for signal versus noise across the different waveform models. Several trends are immediately apparent. First, single-mode models (\(\DSone\), \(\fundamental\)) are strongly disfavored relative to multi-mode descriptions with natural log Bayes factors $\Delta \ln \mathrm{BF}^S_N \gtrsim 130$; together and similarly lower maximum likelihoods. Second, both the $\DStwo$ and $\overtone$ models yield a dramatic fit improvement with respect to their single-mode versions. Notably, the \(\DStwo\) yields a slightly larger Bayes factor and maximum likelihood than $\mathrm{Kerr}_{221}$. In other words, the additional freedom of the agnostic model to fit the data without enforcing Kerr relations between frequencies and damping times is not killed by its larger Ockham Penalty.

The addition of a third damped sinusoid in both models leads to a significantly different increment of the log likelihood and, most importantly, of the Bayes factor. In particular, the maximum log-likelihood is increased only by 3 units in the case of the DS models, in contrast with the 13 units improvement in the Kerr case. This is consistent with what we discussed in our previous section: due to its larger freedom, the $\DStwo$ model was already able to capture portions of the signal that the $\overtone$ model could not capture. Consequently, the addition of a third mode in the $\mathrm{Kerr}_{222}$ model now leads to a dramatically better recovery of the signal as compared to the $\mathrm{Kerr}_{221}$. In both cases, the maximum log likelihood saturates at a value $\max(\ln \mathcal{L}^S_N)\simeq 5190$. We note that the fact that the Kerr models achieve marginally larger values may seem contradictory, given their reduced freedom to fit the data. However, we attribute this to the fact that the sampler is not designed to find the true maximum likelihood value but to make the Bayesian evidence converge.

For both the Kerr and DS models, the Bayesian evidence saturates and $\ln \mathrm{BF}^S_N\simeq 5160$ when 3 \acp{qnm} are included. Adding further sinusoids or overtones does not lead to a gain in \(\max(\ln \mathcal{L}^S_N)\), confirming that a fundamental plus two overtones suffices to capture the signal content at this \ac{snr}. 

The above result indicates that even in the case of an extremely loud ringdown signal, agnostic spectroscopy cannot discriminate between a black-hole that satisfies the no-hair theorem emitting through 3 \acp{qnm} and an alternative object emitting through 3 or 2 damped sinusoids, therefore preventing a statistically significant confirmation of the no-hair theorem, confirming the results in \citep{CalderonBustillo:2020rmh}. {These results seem consistent with the fact that a recent similar analysis on the signal GW250114 yielded a Bayes factor of only $\simeq 3$ in favour of the no-hair theorem (cite) \footnote{While this is not explicitly quoted in the quoted reference, this result can be roughly inferred from the ratio between the prior and posterior distributions in their Figure 4} \footnote{We note that the hair model employed in this work is sligthly more restrictive than ours, as deviations from the expected \ac{qnm} properties are parametrised as fractional deviations from the values imposed by the no-hair theorem.}.}

As expected, the IM-R$_{22}$ model provides the optimal balance between flexibility and physical consistency, with $\ln \mathrm{BF}^S_N \simeq 5167$ and $\max(\ln \mathcal{L}^S_N) \simeq 5190$. This represents the highest Bayesian evidence factor among all models, exceeding both semi-agnostic damped-sinusoid and Kerr-only by seven units. This is, the Kerr nature of the remnant and its compliance with the no-hair theorem can be established with a natural log Bayes Factor of 7 or, equivalently, to the $\gtrsim 99.9\%$ level.

    \subsection{Non-orthogonality of \ac{qnm}}

Previously, we showed in Figures \ref{fig:nested_plot-1} and \ref{fig:tau_plot} how the inferred properties of the QNMs depend strongly on the choice of ringdown model, for the case of a signal consistent with GW150914. In this section, we do the same, but for the case of the simulated signal we have just discussed, which has a ringdown SNR of 100. 

\begin{figure}
    \centering        
    \includegraphics[width=0.49\textwidth]{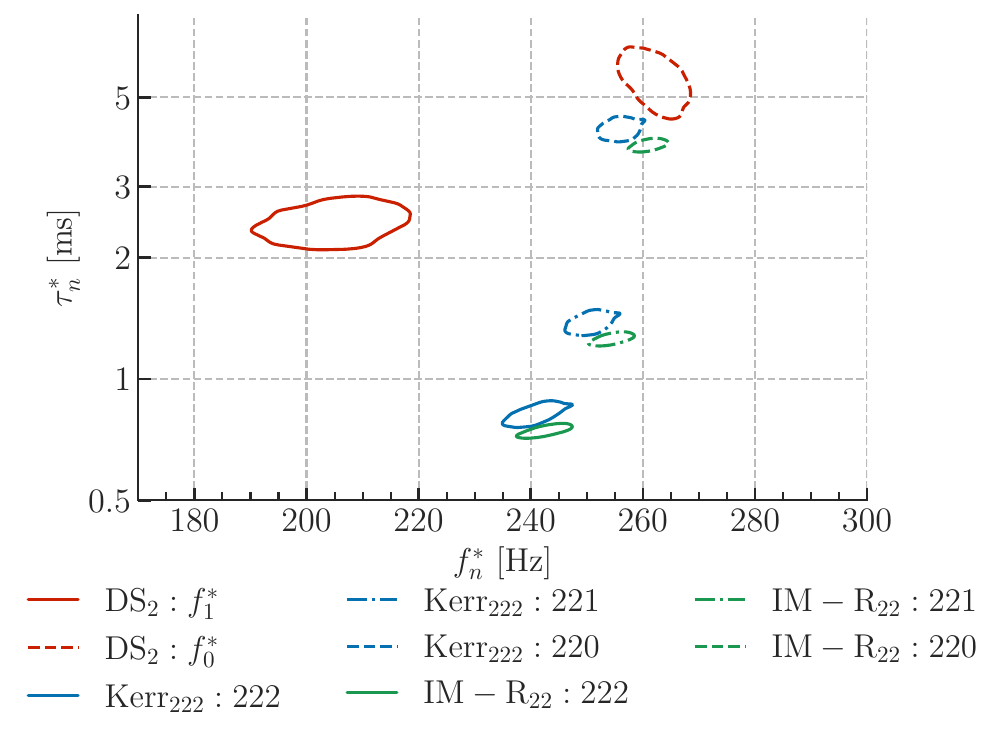}
    \caption{Comparison of QNM frequencies $f_n^\ast$ and damping times $\tau_n^\ast$ inferred for a high-SNR signal at $\tstartgeo=0M$ of an aligned-spin signal restricted to its quadrupole mode under different ringdown models. For the Kerr and IM-R cases, the $(2,2,0)$, $(2,2,1)$, and $(2,2,2)$ mode frequencies and damping times are obtained from the final mass and spin inferred with the $\mathrm{Kerr}_{222}$ and $\mathrm{IM\mbox{-}R}_{22}$ models using the fitting relations of~\citet{Berti:2006wq}. The $y$-axis uses a symmetric logarithmic scale to place both short- and long-lived modes on equal footing.}
\label{fig:high-snr}
\end{figure}

Figure~\ref{fig:high-snr} shows the frequencies and damping times of the inferred QNM content of the signal according to our different ringdown models. In this high-SNR regime ($\mathrm{SNR}=100$), the differences between models become especially clear. Analyses performed with the \(\mathrm{Kerr}_{222}\) and $\DStwo$ models---which are equally preferred by the data---lead to dramatically different conclusions. Both results, moreover, are also inconsistent with those obtained from the fully informed IM-R analysis. In particular, the \(\mathrm{Kerr}_{222}\) model differs from IM-R, likely because weak, individually unresolved \acp{qnm} are effectively absorbed into the limited set of modes available to Kerr-like models. The IM-R framework, by contrast, not only includes the full \acp{qnm} spectrum but also imposes quasi-spherical binary merger motivated priors on their parameters, thereby recovering the correct mode combination. We note that a Kerr model implementing such individually unresolved modes would be statistically disfavoured due to its increased parameter space.
We conjecture that these apparent inconsistencies are likely behind the recent results reported for the signal GW231123~\citep{LIGOScientific:2025rsn}, where semi-agnostic Kerr models infer a \ac{qnm} content at odds with that predicted by full \ac{imr} analyses. Finally, the larger freedom of the $\DStwo$ model produces a mode combination entirely inconsistent with both \(\mathrm{Kerr}_{222}\) and \(\mathrm{IM-R}_{22}\). While the higher frequency $f_0^\ast$ of the longer-lived mode is consistent across models, their damping-time estimates differ. The discrepancies are even more pronounced for the shorter-lived mode, where both the inferred frequencies and damping times are inconsistent.  
        
\section{Conclusion}\label{sec:conclusion}

We have revisited the problem of analyzing black-hole ringdowns by presenting an enhanced version of the method presented in \citet{CalderonBustillo:2020rmh}, where post-merger portions of \ac{imr} waveforms are used to model ringdown signals. Our method enables the analysis of black-hole ringdowns in a way that a) amplitudes and phases of quasinormal modes are naturally constrained to those allowed by given pre-merger dynamics, b) avoids the number of parameters to grow unphysically as modes and overtones are added, c), although not really exploited in this paper, allows for placing priors in terms of the binary parameters. We have improved the approach of \citet{CalderonBustillo:2020rmh} in two main ways. First, we address the treatment of abruptly starting signals through the so-called gating \& inpainting technique. Second, while \citet{CalderonBustillo:2020rmh} modeled the ringdown signal through the post-peak portion of the IMRPhenomPv2 model -- which is restricted to the quadrupole mode -- ours allows for the usage of any waveform model, in particular those including higher-order modes. Finally, our method also trivially permits the usage of ringdown models consisting of collections of \acp{qnm} -- both constrained to satisfy the no-hair theorem and allowing for deviations from it -- allowing us to perform classical (semi-)agnostic spectroscopy.

Applying our formalism to GW150914, we further confirm three main results. First, a classical \acp{qnm} model $(2,2,1)$ -- namely $\overtone$ -- leads to significantly better maximum likelihood (i.e., a better fit) to the data $\Delta {\ln\cal{L}}^{\rm Kerr_{221}}_{\rm Kerr_{220}} \simeq 6$ than that restricting to the $(2,2,0)$ -- which we call $\fundamental$ -- near the signal peak. Indeed, such a maximum log-likelihood is comparable to that obtained when using the post-peak portion of \nrsur, which we denote by IM-R and contains all \acp{qnm}. However, the large parameter space spanned by the $\overtone$ model leads to only a small evidence gain $\Delta \ln \rm{BF}^{\rm Kerr_{221}}_{\rm Kerr_{220}} = 1.4$ with respect to the $\fundamental$ model, translating into weak evidence for the agnostic detection of an overtone, consistently with several previous studies. We note, however, that the fact that IM-R is strongly preferred indicates that the signal must contain modes beyond the fundamental one. Second, using classical spectroscopy, we find that the overtone model is only weakly preferred with respect to a model violating the no-hair theorem, preventing a strong confirmation.
In contrast, using our IM-R model as the reference no-hair model \textit{we confirm the Kerr nature of the remnant to $99\%$ probability}, in accordance with \citet{CalderonBustillo:2020rmh}. Finally, we find no evidence for modes beyond the quadrupole. Next, we have shown that our results are consistent with those obtained on a zero-noise injection consistent with GW150914. Moreover, by analyzing a loud numerically simulated signal consistent with GW150914, we have shown that classical spectroscopic analysis based on the detection of overtones cannot provide a strong confirmation of the no-hair theorem, once again confirming the prediction from \citet{CalderonBustillo:2020rmh}. We have argued that this is consistent with the weak preference for the no-hair theorem found in the recently loud ringdown signal GW250114.

Finally, by analyzing the two mentioned signals, we have explicitly shown the limitations on agnostic black hole spectroscopy imposed by the non-orthogonality of \acp{qnm}. First, we have explicitly shown that the inference of the \ac{qnm} content of the signal is dramatically impacted by the choice of \ac{qnm} model. In particular, we have shown how free damped sinusoid models and Kerr \ac{qnm} models can yield mutually inconsistent inferences, despite being equally supported by the data. Moreover, we have shown that agnostic Kerr QNM models can lead to an inference of the QNM content of the signal that is inconsistent with that predicted by full inspiral-merger-ringdown models, as it has been recently found for the signal GW231123. We have argued that this result is, however, expected. Both models effectively make use of a different number of \acp{qnm}, also effectively imposing dramatically different priors on their properties. In this situation, weak, individually unresolved modes present in the data can be correctly inferred by complete IMR models and by our ringdown-restricted IM-R model, while a Kerr QNM model with a restricted number of modes will simply try to fit such a combination through a ``biased'' mode combination. Moreover, Kerr models that implement such weak unresolved modes would be statistically disfavoured due to the increased Ockham penalty they would incur.  

All in all, our work demonstrates some of the limits of the classical black-hole spectroscopy paradigm and highlights the need for and power of analyses implementing knowledge of the process preceding the formation of the remnant compact object.

\section*{Acknowledgements}\label{sec:acknowledgements}
The authors thank {Elisa Maggio, B. Sathyaprakash, and Gregorio Carullo} for their comments and valuable suggestions, which greatly improved this paper. KC sincerely thanks J. Rzeznik, Isak Danielson, and Mahmood for inspiring him throughout the development of this work. He also thanks the NY Times for providing ideas about better representing the data and ChatGPT for helping improve Figure~\ref{fig:compare-results} by adding iso-central frequency and iso-damping time contours. He acknowledges the generous support provided through NSF grant numbers PHY-2207638, AST-2307147, PHY-2308886, and PHY-2309064. JCB is supported by the Ramon y Cajal Fellowship RYC2022-036203-I and the Grant PID2024-160643NB-I00 of the Spanish Ministry of Science, Innovation and Universities; and by the Grant ED431F 2025/04 of the Galician CONSELLERIA DE EDUCACION, CIENCIA, UNIVERSIDADES E FORMACION PROFESIONAL. JCB also acknowledges support by the programme HORIZON-MSCA2021-SE-01 Grant No. NewFunFiCO-101086251. IGFAE is supported by the Ayuda Maria de Maeztu CEX2023-001318-M funded by MICIU/AEI /10.13039/501100011033.
This research has made use of data, software and/or web tools obtained from the Gravitational Wave Open Science Center (https://www.gw-openscience.org),
a service of LIGO Laboratory, the LIGO Scientific Collaboration, the Virgo Collaboration, and KAGRA. This
material is based upon work supported by NSF’s LIGO Laboratory, which is a major facility fully funded by the
National Science Foundation. LIGO Laboratory and Advanced LIGO are funded by the United States National Science Foundation (NSF) as well as the Science and Technology Facilities Council (STFC) of the United Kingdom, the Max-Planck-Society (MPS), and the State of Niedersachsen/Germany for support of the construction of Advanced LIGO and the construction and operation of the GEO600 detector. Additional support for Advanced LIGO was provided by the Australian Research Council. Virgo is funded through the European Gravitational Observatory (EGO), by the French Centre National de Recherche Scientifique (CNRS), the Italian Istituto Nazionale di Fisica Nucleare (INFN), and the Dutch Nikhef, with contributions by institutions from Belgium, Germany, Greece, Hungary, Ireland, Japan, Monaco, Poland, Portugal, Spain. KAGRA is supported by the Ministry of Education, Culture, Sports, Science and Technology (MEXT), Japan Society for the Promotion of Science (JSPS) in Japan; National Research Foundation (NRF) and the Ministry of Science and ICT (MSIT) in Korea; Academia Sinica (AS) and National Science and Technology Council (NSTC) in Taiwan. 
Finally, the authors acknowledge the Gwave (PSU) cluster's use for computational and numerical work, which was instrumental in conducting the research presented in this paper. 

\bibliography{sample631}{}

\end{document}